\documentclass[acmsmall,screen]{acmart}
\AtBeginDocument{%
  }

\setcopyright{none}
\makeatletter
\renewcommand\@formatdoi[1]{\ignorespaces}
\makeatother
\renewcommand{\footnotetextcopyrightpermission}[1]{}
\usepackage{pifont}
\usepackage{enumitem}
\usepackage{makecell}
\usepackage{mathtools}
\usepackage{multirow}
\usepackage{array}
\usepackage{colortbl}
\usepackage{xcolor}
\usepackage{subfigure}
\usepackage{tcolorbox}
\usepackage{wrapfig}
\usepackage{float}
\usepackage{pict2e}
\usepackage{graphicx}
\usepackage{units}
\usepackage{tikz}
\usepackage{pifont}

\newcolumntype{L}[1]{>{\raggedright\let\newline\\\arraybackslash\hspace{0pt}}m{#1}}
\newcolumntype{C}[1]{>{\centering\let\newline\\\arraybackslash\hspace{0pt}}m{#1}}
\newcolumntype{R}[1]{>{\raggedleft\let\newline\\\arraybackslash\hspace{0pt}}m{#1}}

\newcommand{\RNum}[1]{\uppercase\expandafter{\romannumeral #1\relax}}
\newcommand{\rNum}[1]{\expandafter{\romannumeral #1\relax}}
\newcommand{\md}[1]{#1}

\definecolor{gL1}{RGB}{65,  217, 170}
\definecolor{gL2}{RGB}{136, 252, 197}
\definecolor{gL3}{RGB}{204, 255, 229}
\definecolor{lgray}{gray}{0.95}

\definecolor{rL1}{RGB}{232, 89, 12}
\definecolor{rL2}{RGB}{255, 146, 43}
\definecolor{rL3}{RGB}{255, 192, 120}

\definecolor{textred}{RGB}{255, 51, 51}
\definecolor{textRed}{RGB}{153, 0, 0}
\definecolor{textgreen}{RGB}{0, 204, 0}

\newcommand{\gou}[0]{\ding{52}}
\newcommand{\cha}[0]{\ding{55}}

\tcbuselibrary{skins}
\newtcolorbox{graybox}{
  enhanced,
  boxrule=0pt,
  frame hidden,
  colback=gray!20,  
  borderline west={3pt}{0pt}{gray!50!black},  
  left=4pt,  
  right=4pt,
  top=0.5pt,
  bottom=0.5pt
}

\begin{document}

\title[Coding in a Bubble?]{Coding in a Bubble? Evaluating LLMs in Resolving Context Adaptation Bugs During Code Adaptation}

\author{Tanghaoran Zhang}
\email{zhangthr@nudt.edu.cn}
\orcid{0000-0001-7241-9730}
\affiliation{%
  \institution{National University of Defense Technology}
  \country{China}
}

\author{Xinjun Mao}
\authornote{Xinjun Mao and Yue Yu are corresponding authors.}
\email{xjmao@nudt.edu.cn}
\orcid{0000-0001-6003-5748}
\affiliation{%
  \institution{National University of Defense Technology}
  \country{China}
}

\author{Shangwen Wang}
\email{wangshangwen13@nudt.edu.cn}
\orcid{0000-0003-1469-2063}
\affiliation{%
  \institution{National University of Defense Technology}
  \country{China}
}

\author{Yuxin Zhao}
\email{yuxinzhao@nudt.edu.cn}
\orcid{0009-0005-0061-9457}
\affiliation{%
  \institution{National University of Defense Technology}
  \country{China}
}

\author{Yao Lu}
\email{luyao08@nudt.edu.cn}
\orcid{0000-0002-3520-5829}
\affiliation{%
  \institution{National University of Defense Technology}
  \country{China}
}

\author{Zezhou Tang}
\email{tangzezhou24@nudt.edu.cn}
\orcid{0009-0002-6902-3270}
\affiliation{%
  \institution{National University of Defense Technology}
  \country{China}
}

\author{Wenyu Xu}
\email{xuwenyu@nudt.edu.cn}
\orcid{0009-0006-2088-4976}
\affiliation{%
  \institution{National University of Defense Technology}
  \country{China}
}

\author{Longfei Sun}
\email{lfsun@nudt.edu.cn}
\orcid{0000-0002-8686-818X}
\affiliation{%
  \institution{National University of Defense Technology}
  \country{China}
}

\author{Changrong Xie}
\email{xiechangrong02@nudt.edu.cn}
\orcid{0009-0000-6739-7100}
\affiliation{%
  \institution{National University of Defense Technology}
  \country{China}
}

\author{Kang Yang}
\email{yangkang@nudt.edu.cn}
\orcid{0000-0003-2313-7141}
\affiliation{%
  \institution{National University of Defense Technology}
  \country{China}
}

\author{Yue Yu}
\email{yuy@pcl.ac.cn}
\authornotemark[1]
\orcid{0000-0002-9865-2212}
\affiliation{%
  \institution{Peng Cheng Laboratory}
  \country{China}
}

\renewcommand{\shortauthors}{Zhang et al.}

\begin{abstract}
Code adaptation is a fundamental but challenging task in software development, requiring developers to modify existing code for new contexts. A key challenge is to resolve \textit{\textbf{Context Adaptation Bugs (CtxBugs)}}, which occurs when code correct in its original context violates constraints in the target environment. Unlike isolated bugs, \textit{CtxBugs} cannot be resolved through local fixes and require cross-context reasoning to identify semantic mismatches. Overlooking them may lead to critical failures in adaptation. Although Large Language Models (LLMs) show great potential in automating code-related tasks, their ability to resolve \textit{CtxBugs} remains a significant and unexplored obstacle to their practical use in code adaptation.

To bridge this gap, we propose \textit{CtxBugGen}, a novel framework for generating \textit{CtxBugs} to evaluate LLMs. Its core idea is to leverage LLMs' tendency to generate plausible but context-free code when contextual constraints are absent.
The framework generates \textit{CtxBugs} through a four-step process to ensure their relevance and validity: (1) Selection of four established context-aware adaptation tasks from the literature, (2) Perturbation via task-specific rules to induce \textit{CtxBugs} from LLMs while ensuring their plausibility,
(3) Generation of candidate variants by prompting LLMs without any context constraints and (4) Identification of valid \textit{CtxBugs} through syntactic differencing and test execution in the target context.
Based on the benchmark constructed by \textit{CtxBugGen}, we conduct an empirical study with four state-of-the-art LLMs. Our results reveal their unsatisfactory performance in \textit{CtxBug} resolution. The best performing LLM, Kimi-K2, achieves 55.93\% on Pass@1 and resolves just 52.47\% of \textit{CtxBugs}. The presence of \textit{CtxBugs} degrades LLMs' adaptation performance by up to 30\%. Failure analysis indicates that LLMs often overlook \textit{CtxBugs} and replicate them in their outputs. This suggests that LLMs overly focus on the local code correctness of the reused code while ignoring its compatibility in the target context.
Our study highlights a critical weakness in LLMs' cross-context reasoning and emphasize the need for new methods to enhance their context awareness for reliable code adaptation.
The replication package for this paper is at \url{https://github.com/ztwater/CtxBugGen}.
\end{abstract}



\received{20 February 2007}
\received[revised]{12 March 2009}
\received[accepted]{5 June 2009}

\maketitle

\section{Introduction}
Code reuse is a widely adopted practice in software development, significantly boosting productivity by allowing developers to leverage existing solutions rather than building everything from scratch~\cite{Brandt2009,Gharehyazie2017,Yang2017,Huang2022}. However, their reused code is inherently designed for its original context and may violate specific requirements or constraints of the target context. Thus, developers must always adapt reused code to fit its new context, ensuring its correctness and quality~\cite{Wu2019,Zhang2019,Zhang2024}.
While developers are alert to pre-existing bugs in reused code, they often overlook  the risk of semantic mismatches. This occurs when code that is bug-free in its original context is integrated into a new context without adequate adaptation. These resulting bugs are defined as \textbf{\textit{Context Adaptation Bugs (CtxBugs)}}~\cite{Mondal2019}. Unlike isolated bugs, resolving \textit{CtxBugs} requires developers to perform cross-context reasoning to align the code's behavior with new context.

\begin{figure*}
    \centering
    \includegraphics[width=\linewidth]{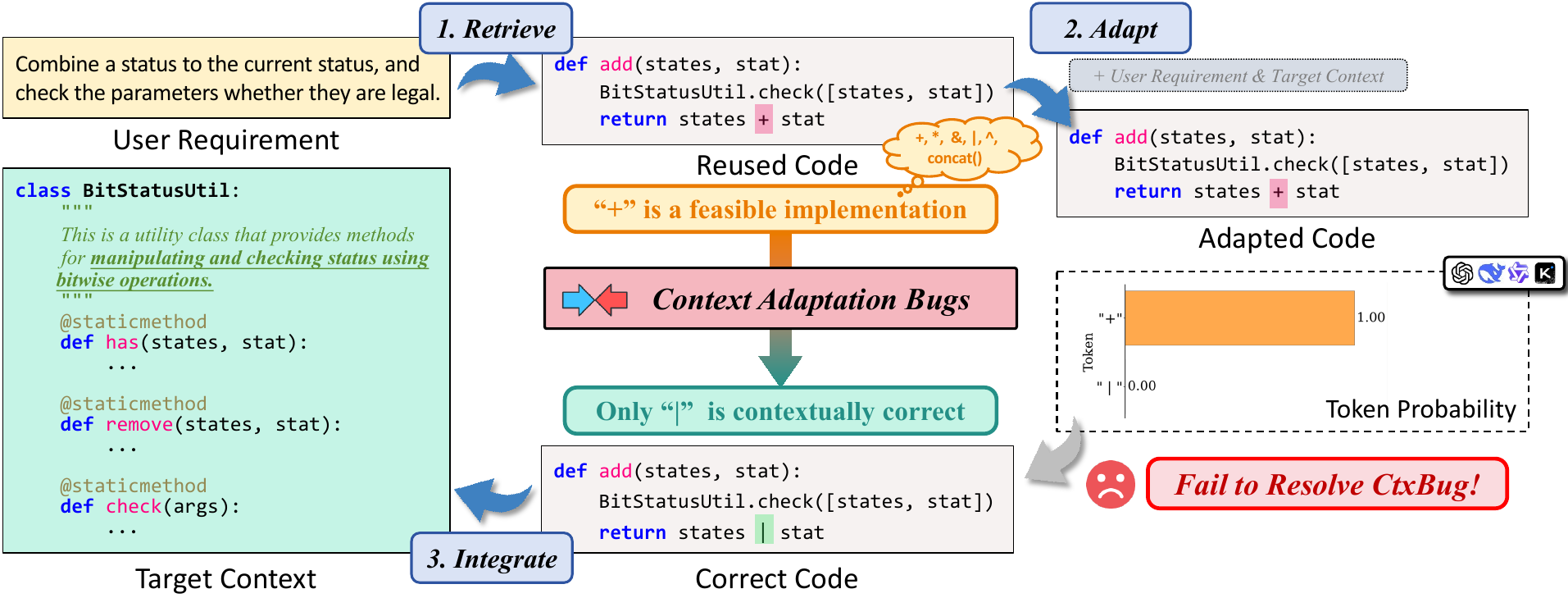}
    \caption{Example of a \textit{CtxBug} in code adaptation. All evaluated state-of-the-art LLMs failed to resolve it.}
    \label{fig:process}
\end{figure*}

Figure~\ref{fig:process} illustrates an example of \textit{CtxBug} in code adaptation. Developer Alice needs to implement a feature for combining bitwise status flags in her utility class. She queries a search engine or an AI assistant with ``How to combine two status values'' and receives a solution using the arithmetic plus operator (``+''). While this numerical addition is a semantically feasible implementation, it introduces a \textit{CtxBug} in Alice's context, where a bitwise OR (``|'') is required to merge flags. This mismatch is subtle, because the reused code compiles and even works for cases like 1 + 2 = 3, but it would fail for others like 1 + 1. This example demonstrates that \textit{CtxBugs} are not intrinsic and isolated bugs in the reused code but semantic mismatches arising from integration, making their resolution critical during the adaptation process.

Recent advancements in large language models (LLMs) have shown great potential in automating various software engineering tasks, including program repair, code optimization, as well as code adaptation~\cite{Xia2023,Gao2024Search-Based,Xu2025Aligning,Zhang2025IorI,Zhang2025AdaptEval}. This capability stems from their powerful prompt-based learning ability~\cite{Brown2020,Liu2021}, which allows them to perform unseen tasks based on instructions provided in a prompt.
However, despite their general applicability, a recent study found that LLMs  perform significantly worse on code adaptation compared to code generation, with a reported performance decrease of 15\%~\cite{Zhang2025IorI}.
We hypothesize that this weakness is relevant to LLMs' limited ability to resolve \textit{CtxBugs} during adaptation. 
To demonstrate this, we apply four state-of-the-art LLMs to the adaptation task presented in Fig.~\ref{fig:process}. In line with the experimental design of Zhang et al.~\cite{Zhang2025IorI}, each LLM was prompted with the requirement, the reused code, and the target context.
Our result indicates that all LLMs failed to resolve the \textit{CtxBug} and expressed high confidence in generating the incorrect operator. Their consistent failures suggest a fundamental limitation in their current cross-context reasoning capability.
Consequently, we argue that rigorously evaluating their effectiveness in resolving \textit{CtxBugs} is a critical prerequisite for their reliable adoption in automated code adaptation.

Despite the significance of \textit{CtxBug} resolution, the lack of dedicated adaptation benchmarks hinders research in this domain.
To address this gap, we propose \textit{CtxBugGen}, a novel framework to generate \textit{CtxBugs} for adaptation evaluation.
Its core insight is that code generated by LLMs with no contextual constraints is inherently general but plausible, making it likely to introduce semantic mismatches for a specific target context, thereby creating \textit{CtxBugs}.
To ensure the relevance and validity of generated \textit{CtxBugs}, \textit{CtxBugGen} employs a four-step pipeline: \textit{\textbf{(1) Adaptation Task Selection}}: We select four context-aware adaptation tasks from current literature to ground our \textit{CtxBugs} in realistic developer needs; \textit{\textbf{(2) Task-Specific Perturbation}}: 
\md{To ensure the plausibility of LLMs' generation, we preserve the overall code structure and perturb minimal but targeted code elements for each adaptation task.}
\textit{\textbf{(3) LLM-Based Variant Generation}}: We instruct LLMs to generate code at the perturbed locations without any contextual constraints. This produces candidates that are semantically correct in isolation but may introduce \textit{CtxBugs}; and \textit{\textbf{(4) CtxBug Identification}}: We combine syntactic differencing with test execution in the target context to identify code with valid \textit{CtxBugs}. 
Finally, we manually validate their quality and obtain a benchmark with 3,683 \textit{CtxBugs}.

Based on our benchmark, we empirically evaluate four state-of-the-art LLMs regarding their \textit{CtxBug} resolution capability. Results demonstrate unsatisfactory performance across all models. The best performing LLM, Kimi-K2, achieves only 55.93\% on Pass@1 and resolves just 52.47\% of \textit{CtxBugs}. Their performance further drops to 36.65\%  Pass@1 and 27.24\% resolution rate when specifically evaluated on the \textit{Functionality Customization} task. Comparative analysis indicates that the presence of \textit{CtxBugs} leads to LLMs' performance degradation of up to 30\%, suggesting their disruptive impact on adaptation. LLMs can even handle isolated bugs more effectively than \textit{CtxBugs} during adaptation. Through failure analysis, we observe that LLMs frequently overlook \textit{CtxBugs} and directly replicate them in their outputs. This suggests that LLMs prioritize the local semantic correctness of code while ignoring their contextual constraints.

This paper makes the following contributions:

\begin{itemize}[leftmargin=*]
    \item We present the first comprehensive study to investigate the effectiveness of LLMs in resolving \textit{CtxBugs}, which is a critical challenge that limits their utilization in code adaptation. 
    \item We propose \textit{CtxBugGen}, a novel framework for effective \textit{CtxBug} generation, and use it to construct the first benchmark for four established adaptation tasks leveraging four LLMs.
    \item Our experimental results reveal the limitations of state-of-the-art LLMs in handling \textit{CtxBugs} and their negative impacts on performance, highlighting the critical need to improve cross-context reasoning in LLMs.
\end{itemize}

\section{Background and Motivation}
\label{sec:background}
This section introduces the background of code adaptation and its recent advances in LLM-based approaches, followed by motivating examples showing the impact of \textit{CtxBugs} on LLMs' adaptation.

\subsection{Code Adaptation}
Code adaptation is a fundamental activity in software development. Although open-source platforms such as Stack Overflow (SO) provide millions of ready-to-use code snippets, integrating them often requires extensive manual efforts~\cite{Wu2019,Zhang2024}. The growing adoption of AI-generated code makes it more challenging, as such code can be difficult to understand and debug~\cite{Barke2023,Li2024AIToolUse}.

Early research, such as Jigsaw~\cite{Cottrell2008}, attempts to automate code adaptation using syntactic similarity. Subsequent studies aim to improve the compilability and reusability of reused code~\cite{Yang2016,Terragni2016CSnippEx}. For instance, Terragni et al. introduce ``APIzation''~\cite{Terragni2021}, which converts code snippets into well-formed, reusable methods, while Zhang et al.~\cite{Zhang2019} leverage historical adaptations to generate reusable templates.
However, these approaches largely overlook the semantic differences across contexts. This gap is identified by Mondal et al.~\cite{Mondal2019}, who introduce the concept of \textit{Context Adaptation Bugs} (\textit{CtxBugs}) that occur when developers clone correct code without proper adaptation. While their study focuses on the evolution of \textit{CtxBugs} in intra-project code clones, it does not explore broader code reuse scenarios, leaving a critical gap in addressing such semantic mismatches during code adaptation.

With the emergence of LLMs, researchers have started to investigate their potential in automating various code adaptation tasks~\cite{Mai2024Code2API,Zhang2025IorI,Wang2025WIRL,Zhang2025AdaptEval}. 
The large-scale parameters and extensive training of LLMs bring them emergent abilities such as prompt-based learning~\cite{Liu2021}. LLMs are allowed to perform unseen tasks which previously require architecture changes and further fine-tuning~\cite{Li2022,Lin2023CCT5}.
One of the most motivational studies is conducted by Zhang et al.~\cite{Zhang2025IorI}, who first investigate LLMs' capabilities in end-to-end adaptation tasks. Their prompt-based approach leverages both the pretrained knowledge and contextual code understanding of LLMs, making it applicable to a wide range of adaptation scenarios. However, their results also reveal that LLMs remain less effective at adaptation tasks than they are at code generation. To further reveal LLMs' strengths and limitations in this task, our study focuses on evaluating their effectiveness in resolving \textit{CtxBugs}.

\subsection{A Motivating Example}
\label{sec:motivation}

\begin{figure*}
    \centering
    \includegraphics[width=\linewidth]{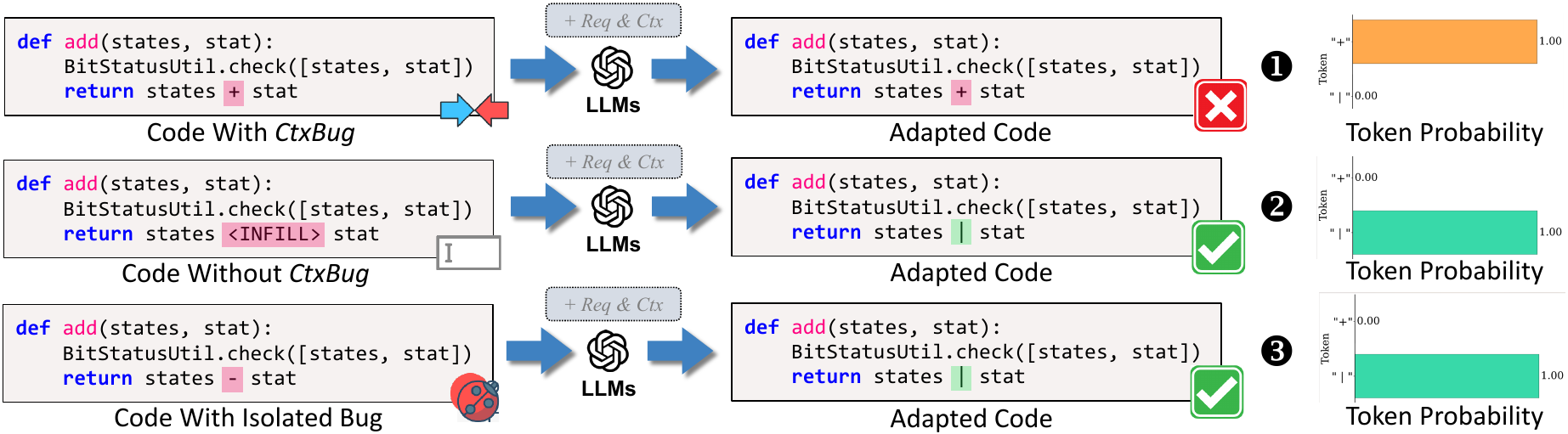}
    \caption{A motivating example to demonstrate how a \textit{CtxBug} can mislead LLMs during adaptation. Bar charts at the right side show LLM-generated token distributions at our concerned position.}
    \label{fig:example}
\end{figure*}

Fig.~\ref{fig:example} illustrates the impact of \textit{CtxBugs} on LLMs' adaptation, which could be misleading and more harmful than isolated bugs. Based on the setup of Fig.~\ref{fig:process}, we employ LLMs for adaptation under three scenarios: code with a \textit{CtxBug}, code without \textit{CtxBugs}, and code with an isolated bug. To simulate these, we replace the bug-inducing operator ``+'' with a placeholder ``\textit{<INFILL>}'' for the bug-free case, and introduce an isolated bug using ``-'' (which violates status combination requirement) in the third case. Then we compare LLMs' outputs and token distributions across all cases.

In Case~\ding{182}, the LLM (e.g., GPT-4o) fails to adapt code with \textit{CtxBug}, assigning 100\% probability to the incorrect token. In contrast, Case~\ding{183} shows that the LLM correctly adapts the code when no \textit{CtxBug} is present. Case~\ding{184} further shows that the LLM can also correctly handle the isolated bug under the same setting. In both correct scenarios, the LLM shows no tendency to generate the incorrect token, while it is misled by the \textit{CtxBug} in the first case.
These results suggest that LLMs are susceptible to \textit{CtxBugs}, likely because \textit{CtxBugs} appear semantically consistent with the local context, such as the method name ``\textit{add}''. 
This hinders LLMs from capturing global context constraints (e.g., bitwise state operations) and leads to their poor performance in code adaptation.
The above observation motivates our study to systematically evaluate LLMs' capabilities in resolving \textit{CtxBugs}.

\section{\textit{CtxBugGen} Framework}

\begin{figure*}
    \centering
    \includegraphics[width=\linewidth]{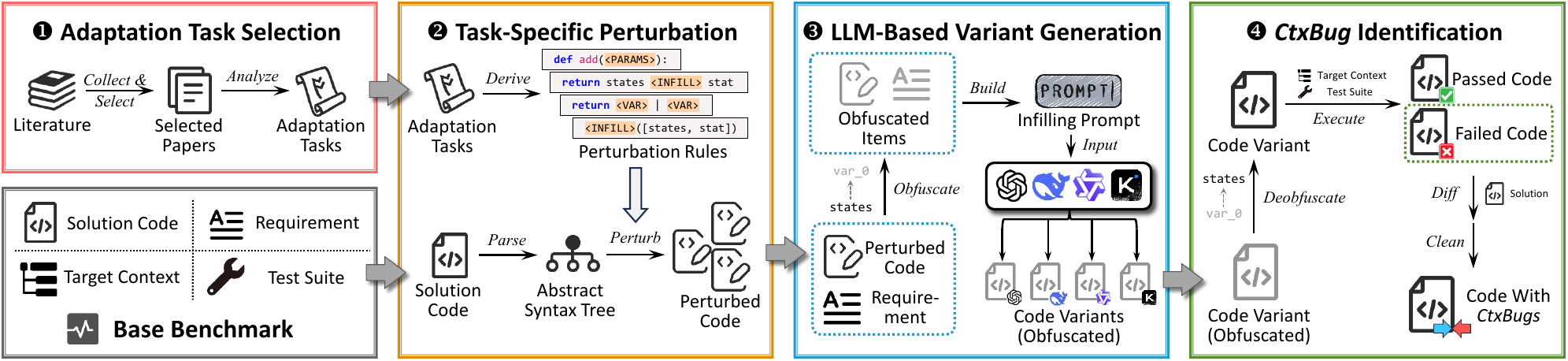}
    \caption{The benchmark construction process through the \textit{CtxBugGen} framework.}
    \label{fig:framework}
\end{figure*}

This section introduces \textit{CtxBugGen}, a framework for constructing evaluation benchmarks containing \textit{CtxBugs}. Its core idea is to leverage LLMs to generate code by deliberately withholding contextual constraints. This elicits semantically plausible but contextually incorrect outputs, thereby introducing \textit{CtxBugs}.
An overview of \textit{CtxBugGen} is shown in Fig.~\ref{fig:framework}. It includes four steps: adaptation task selection (Section~\ref{sec:s0-task-select}), task-specific perturbation (Section~\ref{sec:s1-perturb}), LLM-based variant generation (Section~\ref{sec:s2-variant-gen}) and \textit{CtxBug} identification (Section~\ref{sec:s3-csc-identify}). Each step is detailed as follows.

\subsection{Adaptation Task Selection}
\label{sec:s0-task-select}
To ensure our benchmark reflects practical adaptation challenges, we systematically analyze the literature to answer a key question:

\textbf{What are the primary code adaptation issues identified in existing studies?}

Following guidelines for mapping studies in software engineering~\cite{Kitchenham2007Guidelines,Petersen2015Guidelines}, we first conduct an explicit literature collection and selection process to retrieve relevant papers. We then extract and analyze all data items related to code adaptation issues through thematic analysis~\cite{Braun2006}, and finally report our results. All data and materials are available in our replication package.

\subsubsection{Literature Collection and Selection}
\label{sec:paper-collect}

\begin{figure*}
    \centering
    \includegraphics[width=\linewidth]{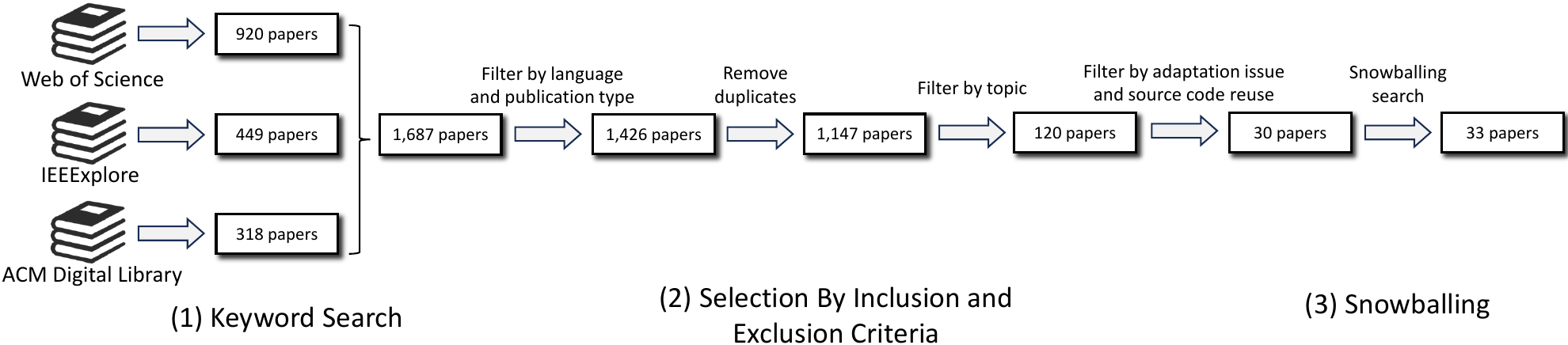}
    \caption{Our literature collection and selection process.}
    \label{fig:lr}
\end{figure*}

As shown in Fig.~\ref{fig:lr}, our literature collection and selection process consists of three steps: (1) keyword search, (2) selection by inclusion and exclusion criteria and (3) snowballing. The detail of each step is as follows. 

\textbf{Keyword Search.} In line with prior surveys ~\cite{Yang2022survey,Yang2024Survey}, we select Web of Science, IEEEXplore and ACM Digital Library as our databases. As we aim to cover papers about code adaptation as comprehensive as possible, we also include \textit{code reuse} and \textit{code integration} along with the most precise term, \textit{code adaptation} in our keywords. To the best of our knowledge, the first related study on source code adaptation was Jigsaw~\cite{Cottrell2008Jigsaw}, which was published in 2008. Hence, the time interval of collection is set from 2008 to 2025, lasting for 17 years. This step yields 1,687 papers in total.

\textbf{Selection By Inclusion and Exclusion Criteria.} This step aims to further evaluate the quality and relevance of our collect papers. Our criteria are shown in Table~\ref{tab:criteria}. The selection is conducted by two researchers with rich experience in software engineering and code adaptation. We first use automatic filters to apply \textit{Exclusion Criteria 1-3}, resulting in 1,147 papers. Then we manually read the title, abstract and keywords of each paper to determine whether they fit our topic according to \textit{Inclusion Criteria 1}. After this step, we narrow down to 120 papers. Subsequently, we look through these papers to identify source code reuse and related adaptation issues (\textit{Inclusion Criteria 2 \& Exclusion Criteria 4}), resulting in 30 papers.

\begin{table}[htbp]
    \scriptsize
    \centering
    \caption{Inclusion and exclusion criteria of our literature selection.}
    \begin{tabular}{cL{345pt}}
        \toprule
        \textbf{Criteria} & \multicolumn{1}{c}{\textbf{Explanation}} \\
        \midrule
        \multirow{2}{*}{Inclusion} 
        & 1. The paper should be related to the topic of code reuse or adaptation. \\
        & 2. The paper should focus on at least one code adaptation issue derived from empirical evidences. \\
        \midrule
        \multirow{4}{*}{Exclusion} 
        & 1. Non-English written paper should be discarded. \\
        & 2. Books, chapters, thesis, or grey publications should be discarded. \\
        & 3. Duplicated papers included in multiple databases or similar studies with different versions should be discarded.\\
        & 4. The paper without including \textit{source code reuse} should be discarded, e.g., component, library or non-code artifact reuse. \\
        \bottomrule
    \end{tabular}
    \label{tab:criteria}
\end{table}

\textbf{Snowballing.} To identify additional relevant studies, we conduct a snowballing search in the last step. By looking through the reference and citation list of above 30 papers, we manually add three papers which are missed by our keyword search. Additionally, we ask another researcher to double-check our final list of papers to ensure that they satisfy our criteria.

\subsubsection{Data Item Extraction and Analysis}
\label{sec:analysis}
\begin{table}[htbp]
    \scriptsize
    \centering
    \caption{Adaptation issues identified by existing code adaptation literature.}
    \begin{tabular}{C{48pt}L{175pt}L{75pt}cc}
        \toprule
        \textbf{Category} & \multicolumn{1}{c}{\textbf{Description}} & \multicolumn{1}{c}{\textbf{Related Studies}} & \textbf{Num} & \textbf{Ctx-Based} \\
        \midrule
        Interface Specification 
        & Mismatches between the \textit{interface} (e.g., signature) of the reused code and its specifications in the target context.
        & \cite{Yang2016,Wang2016Hunter,Terragni2021,Mai2024Code2API,Zhang2024,Zhang2025IorI} 
        & 6 & \gou \\
        \midrule
        Functionality Customization 
        & Mismatches between the \textit{internal logic} (e.g., conditionals) of reused code and its desired behavior in the target context.
        & \cite{Doerner2014Euklas,Yang2016,Terragni2016CSnippEx,Wu2019,Zhang2019,Shen2021VarInAPI,Yang2023Jupyter,Reid2023NCQ,Zhang2024,Zhang2025IorI} & 10 & \gou  \\
        \midrule
        Identifier Reference 
        & Incorrect \textit{identifier references} to the local (e.g., local variables) or contextual entities (e.g., user-defined methods).
        & \cite{Jablonski2007CReN,Cottrell2008Jigsaw,Wightman2012,Holmes2012,Doerner2014Euklas,Yang2016,Terragni2016CSnippEx,Allamanis2017SmartPaste,Campos2019,Zhang2019,Reid2020,Shen2021VarInAPI,Liu2022AdaptivePaste,Zhang2024,Zhang2025IorI,Wang2025WIRL} & 16 & \gou \\
        \midrule
        Dependency Constraint & Mismatches and misuses of \textit{dependencies} that are incompatible with the target context.
        & \cite{Nguyen2010ApiAdapt,Holmes2012,Doerner2014Euklas,Armaly2014ExecReplay,Terragni2016CSnippEx,Zhang2018ExampleCheck,Reid2020,Ragkhitwetsagul2021,Huang2023PCRChain,Zhang2024,Zhang2025IorI} & 11 & \gou \\
        \midrule
        \multirow{2.4}{*}{Miscellaneous} 
        & Reuse code without providing \textit{proper attribution} to its origin.  & \cite{Sojer2011,An2017,Lotter2018,Ragkhitwetsagul2021,Colombo2025} & 5 & \cha \\
        \cmidrule{2-5}
        & \textit{Vulnerabilities} could spread through code reuse if not resolved. & \cite{Fischer2017SP,Reid2022Orphan,Yang2023Jupyter} & 3 & \cha \\
        \bottomrule
    \end{tabular}
    \label{tab:lr-issue}
\end{table}

In this step, we extract the essential data items from our selected papers, which helps us summarize current adaptation issues. Two researchers annotate data items following the principles of thematic analysis~\cite{Braun2006}. It includes two phases: (1) independently code each data item with its representative information, e.g., what property of source code is mentioned; and (2) work together to discuss about initial codes and group them into themes with descriptive names. To avoid bias, we invite another researcher to reassign the established themes to each item for cross validation. Its Cohen's Kappa~\cite{Cohen1960} is 0.932, which indicates an almost perfect agreement between annotators. Results are shown in Table~\ref{tab:lr-issue}, including five categories (themes) of adaptation issues. Their detailed explanations are in the following.

\textbf{Interface Specification} issues refer to mismatches between the interface of the reused code and its required specifications in the target context. Specifically, developers often reuse code from sources like Stack Overflow (SO), where snippets may lack proper method signatures or contain dangling code.  
Hence, they have to perform a specific type of adaptation, which is called ``APIzation''~\cite{Terragni2021,Mai2024Code2API}, to encapsulate such code into well-formed methods with appropriate parameters and return statements. Besides, developers may also modify the parameter list by parameterizing some local variables or hard-coded values to improve reusability~\cite{Zhang2024}. 

\textbf{Functionality Customization} issues refer to mismatches between the internal logic of reused code and its desired behavior in the target context.
Prior studies find that code snippets from platforms like SO are often illustrative and may be semantically incomplete or not directly usable~\cite{Yang2016,Terragni2016CSnippEx,Zhang2019}. Hence, developers must frequently customize their functionality~\cite{Wu2019,Zhang2019}. Even when code is complete and executable, it may still require adaptations to satisfy specific needs, e.g., extracting relevant logic from code with tangled features~\cite{Zhang2024}. This task covers adaptations of a wide range of code elements, including constants, types, conditionals, and API usage~\cite{Zhang2019,Shen2021VarInAPI}.

\textbf{Identifier Reference} issues involve incorrect references to local or contextual entities and fall into two sub-categories: (1) misuse of bounded identifiers (already defined in the target context), and (2) misuse of free identifiers (instantiated within the current scope).
For bounded identifiers, previous studies identify a common adaptation pattern called \textit{Code Wiring}, where developers replace undeclared or conflicting variables with contextually appropriate ones~\cite{Zhang2024,Wang2025WIRL}. Similar adaptations are also applied to method identifiers to align with existing implementations.
For free identifiers, adaptations often address variable misuse issues, which occur when integrating code with duplicate variable names or confusing variables of similar types and purposes~\cite{Allamanis2017SmartPaste,Liu2022AdaptivePaste}. 

\textbf{Dependency Constraint} issues refer to mismatches and misuses of dependencies that are incompatible with the target context, including two main types: (1) missing or inconsistent imports: Code snippets from sources like SO often lack necessary imports or use incorrect library names or aliases, requiring developers to align the dependencies with the target context~\cite{Terragni2016CSnippEx,Reid2020,Huang2023PCRChain,Mai2024Code2API}
(2) version and compatibility issues: Rapid evolution of third-party libraries often leads to altered, replaced, or removed APIs. Incompatible or deprecated APIs can cause failures when reused code and the target context depend on different library versions~\cite{Nguyen2010ApiAdapt}. Reusing outdated snippets or those containing API violations~\cite{Zhang2018ExampleCheck,Ragkhitwetsagul2021} without appropriate adaptation may also introduces risks.

We also identify two \textbf{Miscellaneous} issues from the literature: (1) license violations, where code is reused without proper attribution to its original source~\cite{Sojer2011,An2017,Lotter2018,Ragkhitwetsagul2021}; and (2) spread of vulnerabilities, where vulnerabilities in reused code may spread to the target system if not detected and resolved during adaptation~\cite{Fischer2017SP,Reid2022Orphan}. 
Although these issues are critically important in practice, they fall beyond the scope of our study on \textit{CtxBugs}. Licensing is primarily a legal concern rather than a semantic one, while vulnerabilities are generally context-agnostic. They reflect inherent flaws in the code, rather than bugs that emerges from integrating code into a new context. Therefore, we focus on the first four context-dependent issues in \textit{CtxBug} generation and LLM evaluation, which constitute our \textbf{\textit{adaptation tasks}}.

\subsection{Task-Specific Perturbation}
\label{sec:s1-perturb}
In this step, we transform solution code from an existing benchmark into perturbed code templates for LLM-based variant generation. Specifically, we first derive ten perturbation rules based on our adaptation tasks. Subsequently, we parse the solution code into an abstract syntax tree (AST) using \textit{tree-sitter-python}. By traversing the AST, we replace the target code elements with their corresponding placeholders according to our derived rules. 

\subsubsection{Selected Base Benchmark} To support \textit{CtxBug} generation, a base benchmark should include the following components: (1) comprehensive requirement specification to describe the intended functionality of reused code, (2) rich code context to simulate the target environment, and (3) high-quality test suites to detect the constraint violation in the target context. To this end, we select a typical context-aware benchmark, ClassEval~\cite{Du2024}, as our base benchmark. Specifically, it is a class-level code generation benchmark widely adopted for evaluating LLMs in code generation~\cite{Chen2025REval} and also adaptation~\cite{Zhang2025IorI}. It includes detailed descriptions and test cases for each class and method, as well as rich intra-class interactions and diverse library usage. Besides, ClassEval also covers a broad range of programming topics that reflect real-world development scenarios. Following Zhang et al.~\cite{Zhang2025IorI}, we focus on integrating function-level code within class-level contexts. It is worth noting that our framework is generalizable and can be applied to other benchmarks.

\subsubsection{Perturbation Rules}
\md{A straightforward approach to generating \textit{CtxBugs} is to mask entire methods and use LLMs to generate code from scratch without contextual constraints. However, this often results in implausible or functionally divergent outputs, e.g., implementing real mail-sending logic instead of a simulation. Besides, developers also prefer reusing near-correct code to avoid costly adaptations in practices~\cite{Wu2019,Zhang2024}, hence our approach preserves the overall code structure and applies minimal perturbations to elements relevant to specific adaptation tasks. Guided by the adaptation taxonomy and syntactic rules from Zhang et al.~\cite{Zhang2019}, we design ten perturbation rules at two granularities: token-level and expression-level. To prevent LLMs from trivially inferring correct values from local context, e.g., deducing a boolean value from a contradictory condition or replicating a preceding arithmetic operation, we perturb all instances of a specific token type or all occurrences of a particular expression in a single perturbation. This allows a freer generation while preserving semantic plausibility. The detailed rules are described below.}

\textbf{Interface Specification.}
This task involves adapting the inputs (parameters) and outputs (return statements) of a method. As these elements are semantically distant and define distinct aspects of the interface, we perturb them individually to induce \textit{CtxBugs}.
It is worth noting that masking occurrences of a specific return statement may also allow LLMs to bypass the reasoning process, e.g., by inferring a \textit{return True} from a visible \textit{return False} statement.
To this end, we mask all return statements simultaneously using explicit ``\textit{<RETURN>}'' placeholders. This not only eliminates LLMs' shortcuts in reasoning but also constrains them to focus exclusively on generating return values. 

\begin{graybox}
    \textbf{[Rule 1 Parameter]:} \textit{Replace the parameter list with a ``<PARAMS>'' placeholder.}
    
    \textbf{[Rule 2 Return Statement]:} \textit{Replace all return statements with ``<RETURN>'' placeholders.}
\end{graybox}

\textbf{Functionality Customization.} 
This task focuses on adapting the internal code logic to aligned with the desired behavior in the context. Prior work identifies that such adaptations commonly involve constants, operations, conditionals, and API calls~\cite{Zhang2019}. Implanting errors to these elements is also an established practice for creating program repair benchmarks, e.g., DebugBench~\cite{Tian2024}, CodeEditorBench~\cite{Guo2024}. Thus, we perturb them individually to induce functionality-related \textit{CtxBugs}. 

Constants represent critical but context-specific values for code logic, e.g., array indices, API options. Since they are defined at the token level, we perturb all instances of a given constant type (e.g., all integers or strings) to induce \textit{CtxBugs} that violate hard-coded contextual constraints.

\begin{graybox}
    \textbf{[Rule 3 Constant]:} \textit{Replace all constants of a specific type with ``<INFILL>'' placeholders.}
\end{graybox}

Operations are critical for controlling program behavior. To induce more \textit{CtxBugs} in their semantics, we perform perturbations at both granularities. 
Firstly, we perturb all token-level operators as they encode the core logic of any operation. 
Secondly, we perturb each right-hand side of assignment statements individually to introduce higher-level semantic differences.

\begin{graybox}
    \textbf{[Rule 4 Operator]:} \textit{Replace all operators with ``<INFILL>'' placeholders.}
    
    \textbf{[Rule 5 Operation]:} \textit{Replace each right-value operation with a ``<INFILL>'' placeholder.}
\end{graybox}

Since token-level perturbations above also cover conditionals and function calls, the following rules focus on their expression-level perturbations. Based on our criteria, we perturb their each occurrence individually. For library API calls, this rule focuses solely on their functional purposes, rather than on environmental constraints. Hence, we do not perturb other library usage simultaneously to restrict LLMs, which is different from \textbf{Rule 10}. Additionally, methods defined in the target contexts are excluded, as bounded identifier usage is specifically addressed in \textbf{Rule 8}.

\begin{graybox}
    \textbf{[Rule 6 Conditional]:} \textit{Replace each conditional with a ``<INFILL>'' placeholder.}
    
    \textbf{[Rule 7 Function Call]:} \textit{Replace each function call with a `<INFILL>' placeholder.}
\end{graybox}

\textbf{Identifier Reference.} This task focuses on whether identifiers are correctly resolved within the target context. Due to the differences of bounded and free identifiers described in Section~\ref{sec:s0-task-select}, we address them separately. For bounded identifiers, their references include attributes, methods and global identifiers. We perturb all their occurrences to prevent LLMs from accessing their contextual information. For free identifiers, we perturb all occurrences of local variable identifiers except their first declaration, in line with the variable misuse task defined in~\cite{Allamanis2017SmartPaste,Liu2022AdaptivePaste}. Since this task specifically targets variables, we use explicit ``\textit{<VAR>}'' placeholders to guide LLMs' generation.

\begin{graybox}
    \textbf{[Rule 8 Bounded Identifier]:} \textit{Replace all bounded identifiers with ``<INFILL>'' placeholders.}

    \textbf{[Rule 9 Free Identifier]:} \textit{Replace all local variable identifiers with ``<VAR>'' placeholders.}
\end{graybox}

\textbf{Dependency Constraint.} This task focuses on resolving incompatible library usage arising from environmental factors. 
To force LLMs to generate feasible alternatives that may lead to \textit{CtxBugs}, we perturb all attributes and API calls from third-party libraries simultaneously.
\begin{graybox}
    \textbf{[Rule 10 Library Usage]:} \textit{Replace all library attributes or API calls with ``<INFILL>'' placeholders.}
\end{graybox}

By applying above rules, we derive 3,527 perturbed code instances from 410 methods in the base benchmark.
While some perturbed elements may be repetitive or overlapping, the goal of this step is to better induce \textit{CtxBugs} rather than ensure uniqueness. Any duplicate variants generated by LLMs will be further filtered during the identification process (Section~\ref{sec:s3-csc-identify}).

\subsection{LLM-Based Variant Generation}
\label{sec:s2-variant-gen}
In this step, we leverage LLMs to generate code variants containing potential \textit{CtxBugs} from our perturbed code. However, since our base benchmark ClassEval was released in 2023, its inclusion in the latest LLMs' training corpora creates a risk of data leakage. A direct consequence is that LLMs may simply output the solution code even when provided only with the perturbed code and no target context. This bypasses their reasoning and results in no \textit{CtxBugs} being generated.

To ensure that LLMs' generation adheres strictly to our requirements and is not influenced by prior exposure to the base benchmark, we employ code obfuscation techniques~\cite{Zhang2025Unseen,Kong2025} to mitigate the potential data leakage issue. Specifically, we rename all the identifiers in the perturbed code and their corresponding mentions in the requirement descriptions, e.g., $state\rightarrow var\_0$. This requires LLMs to complete the code according to requirement rather than relying on memorized data. This obfuscation step does not harm our purpose, as the obfuscated code is only textually different while preserving the original semantics, ensuring its validity during LLMs' inference phase.

Subsequently, we construct a set of code infilling prompts, each containing only the requirement description and a perturbed code snippet, to guide LLMs in generating code variants without exposing any contextual constraints.
For the \textit{Dependency Constraint} task, we include an additional instruction to explicitly restrict the use of correct libraries: ``\textit{However, you are not allowed to use the following libraries: [LIB\_DEPS]}'' where \textit{LIB\_DEPS} denotes all third-party libraries used in the solution. To ensure the robustness of our framework, we employ four different LLMs for \textit{CtxBug} generation, which are the same models used in our subsequent evaluation (Section~\ref{sec:studied-llms}).

\subsection{\textit{CtxBug} Identification}
\label{sec:s3-csc-identify}

\begin{wrapfigure}[10]{R}{0.49\textwidth}
    \centering
    \includegraphics[width=0.47\columnwidth]{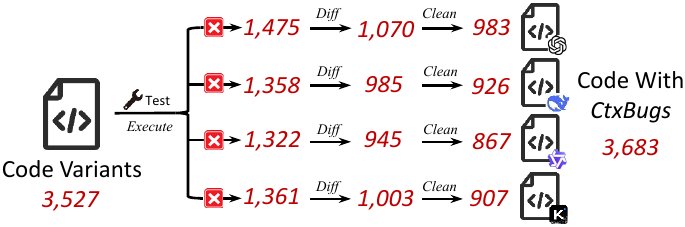}
    \caption{Details of \textit{CtxBug} identification. The numbers represent the remaining cases after each filtering step.}
    \label{fig:detail}
\end{wrapfigure}

This step employs a hybrid approach to identify \textit{CtxBugs} from LLM-generated code variants. Since \textit{CtxBugs} should be contextually incorrect, we first deobfuscate all variants to restore original identifier names and test them within the target context. However, test execution alone is insufficient, because it cannot determine whether a failure stems from an incorrect completion at the perturbed location or from irrelevant changes introduced elsewhere by LLMs. Relying solely on execution could result in \textit{CtxBugs} that do not align with the intended adaptation task. 

To address this, we complement execution-based identification with syntactic differencing. This technique ensures that the generated code differs from the solution only at the intentionally perturbed locations. The algorithm for this process is described as follows. We first parse both the LLM-generated variant and solution code into ASTs. We then use GumTree~\cite{Falleri2014}, a widely-used code differencing tool, to obtain AST node matches and an edit script between two ASTs. For each perturbed node in the solution code (obtained in Section~\ref{sec:s1-perturb}), we identify its corresponding matching node in the variant. A \textit{CtxBug} is considered valid if either: (1) a matching node is found and its textual content differs from the original, or (2) the edit script indicates that the perturbed node is deleted.
If all perturbed nodes are found to be textually identical in the variant, the case is filtered out as it does not contain a \textit{CtxBug}. This process confirms that any functional incorrectness arises from a genuine \textit{CtxBug} at the intended location, rather than from arbitrary or extraneous modifications.

However, our hybrid identification approach cannot handle empty outputs. Furthermore, different perturbation rules may generate identical code variants containing the same \textit{CtxBug}. To prevent these issues from affecting our evaluation, we perform a data cleaning step to filter out repetitive or empty outputs. Detailed statistics for the identification process for each LLM are provided in Fig.~\ref{fig:detail}.

To ensure quality, we further conduct manual validation for 348 snippets containing \textit{CtxBugs} by employing a stratified sampling (with 95\% CI and MOE=5\%). Two annotators independently evaluate each sample against our criteria for \textit{CtxBugs}: (1) whether the code is plausible regarding the requirement, (2) whether it is incorrect in the target context, and (3) whether generated \textit{CtxBugs} are located at the intended locations. Since criterion 2 is validated by test execution, our validation focuses on criteria 1 and 3. For any instance failing them, the annotators document the specific reason. Our validation results demonstrate the high quality of the benchmark. 90.80\% of the sampled \textit{CtxBugs} satisfy all criteria. Only 30 cases (8.62\%) contain isolated bugs and 4 cases (1.15\%) have the location issue. The inter-annotator agreement is nearly perfect, with Cohen's Kappa values~\cite{Cohen1960} of 0.866 (criterion 1) and 0.886 (criterion 3). The statistics of our benchmark (\textit{w/ CtxBugs}) are presented in Table~\ref{tab:dataset}. The full benchmark and our annotated data are available in the replication package. 

\section{Experimental Setup}
This section describes our experimental setup to evaluate the effectiveness of LLMs in resolving \textit{CtxBugs}. Specifically, we structure our evaluation through the following three research questions. 

\begin{figure*}
    \centering
    \includegraphics[width=\linewidth]{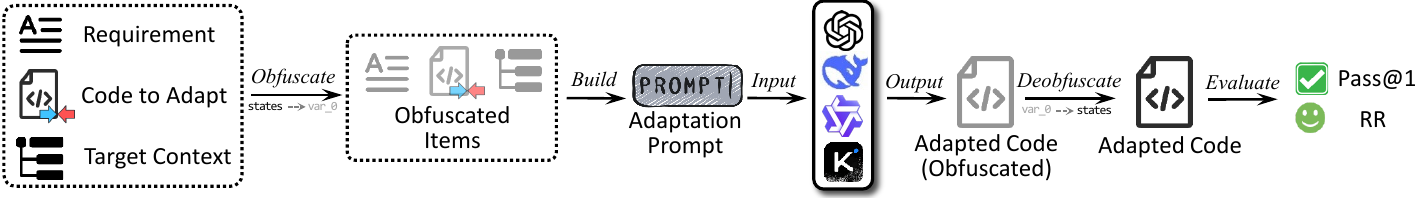}
    \caption{The evaluation process for LLMs' adaptation.}
    \label{fig:evaluation}
\end{figure*}

\textbf{RQ1: How effective are LLMs at resolving \textit{CtxBugs} during code adaptation?} This RQ aims to benchmark and compare the performance of state-of-the-art LLMs in resolving \textit{CtxBugs} generated from \textit{CtxBugGen}. Besides, it also investigates whether their effectiveness is impacted by the source model that introduces \textit{CtxBugs}.

\textbf{RQ2: To what extent do \textit{CtxBugs} impact code adaptation performance of LLMs?} This RQ compares LLMs' performance across three settings: adapting code with \textit{CtxBugs}, code without \textit{CtxBugs} and code with isolated bugs. The comparison aims to quantify the impact of \textit{CtxBugs} and highlight their differences from isolated bugs during adaptation.

\textbf{RQ3: What are the limitations of LLMs in resolving \textit{CtxBugs} during adaptation?} This RQ involves a qualitative analysis of LLMs' failures when resolving \textit{CtxBugs}. We investigate their erroneous handling actions and common \textit{CtxBug} types in different adaptation tasks. The findings aim to provide insights for improving LLMs' capabilities and their practical use in code adaptation.

Our evaluation process is illustrated in Fig.~\ref{fig:evaluation}. To mitigate data leakage issue, we employ a class-level code obfuscation for LLMs' inputs. Then we evaluate four state-of-the-art LLMs with a designed adaptation prompt through execution and syntactic metrics. The details are as follows.

\subsection{Baselines}

\begin{figure*}
    \centering
    \includegraphics[width=\linewidth]{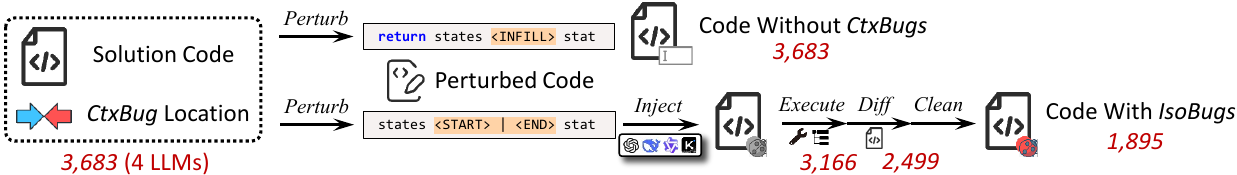}
    \caption{The data construction process for baselines.}
    \label{fig:baselines}
\end{figure*}

To measure the impact of \textit{CtxBugs} on LLMs' adaptation, we design two baselines: (1) \textit{Code Without CtxBugs} and (2) \textit{Code With Isolated Bugs (IsoBugs)}, whose construction processes are illustrated in Fig.~\ref{fig:baselines}. 
For \textit{Code Without CtxBugs}, we directly replace all \textit{CtxBug} locations with ``\textit{<INFILL>}'' placeholders. This represents the setting without distractions from their semantics, forcing LLMs to merely reason with context. This approach aligns with the established program repair practices~\cite{Xia2022AlphaRepair}. However, the placeholders may leak positional information by signaling to LLMs which locations require completion. 
To address this issue, we introduce a second baseline, \textit{Code With IsoBugs}. Its construction leverages LLMs' ability to implant bugs~\cite{Tian2024,Guo2024}. Specifically, we first enclose each code span where \textit{CtxBug} is located within ``\textit{<START>}'' and ``\textit{<END>}'' tokens in the solution code. This design enables a direct paired comparison of bug resolution performance at identical locations. We then prompt LLMs to introduce isolated bugs within the marked span. The generated candidates undergo the same identification process as \textit{CtxBugs}, including test execution, syntactic differencing, and data cleaning, as their only difference lies in local correctness. 
To ensure quality, we also conduct manual validation by sampling 320 snippets (with 95\% CI, MOE=5\%). Two annotators independently evaluate each sample against two criteria: (1) whether the code contains a bug in isolation, and (2) whether generated \textit{IsoBugs} are located at the target location. Our validation results show that 92.81\% (297/320) of the generated \textit{IsoBugs} are both locally incorrect and correctly located, with Cohen's Kappa values~\cite{Cohen1960} of 0.832 (criterion 1) and 1.0 (criterion 2), respectively, indicating nearly perfect agreement. The detailed statistics are listed in Table~\ref{tab:dataset}. The baseline benchmarks and our annotated data are available in the replication package.

\begin{table}[htbp]
    \scriptsize
    \centering
    \caption{The statistics of generated benchmarks. \textit{Code with} and \textit{without CtxBugs} have identical sizes, as they differ only at the perturbed locations. \textit{Code with IsoBugs} undergoes an extra generation and filtration process and thus has a smaller size. Values in parentheses indicate the total number of bugs for each category.
    }
    \begin{tabular}{cc|C{45pt}C{45pt}C{45pt}C{45pt}|C{45pt}}
        \toprule
        \textbf{Model} & \textbf{Setting}
        & \textbf{Interface} & \textbf{Functionality}
        & \textbf{Identifier} & \textbf{Dependency}
        & \textbf{All Tasks} \\
        \midrule
        & \textit{w/(o) CtxBugs} 
        & \textbf{259}\ \,(320) & \textbf{379}\ \,(497) 
        & \textbf{297}\ \,(835) & \textbf{48}\ \,(109) 
        & \textbf{983}\ \,(1,761) \\
        \rowcolor{gray!25} \cellcolor{white} \multirow{-2}{*}{GPT-4o}
        & \textit{w/ IsoBugs} 
        & \textbf{112}\ \,(161) & \textbf{291}\ \,(389) 
        & \textbf{208}\ \,(583) & \textbf{45}\ \,(106) 
        & \textbf{656}\ \,(1,239) \\
        \midrule
        \multirow{3.4}{*}{DS-V3}
        & \textit{w/(o) CtxBugs} 
        & \textbf{242}\ \,(286) & \textbf{366}\ \,(491) 
        & \textbf{271}\ \,(785) & \textbf{47}\ \,(125) 
        & \textbf{926}\ \,(1,687) \\
        \rowcolor{gray!25} \cellcolor{white} \multirow{-2}{*}{DS-V3}
        & \textit{w/ IsoBugs} 
        & \textbf{100}\ \,(133) & \textbf{238}\ \,(339) 
        & \textbf{132}\ \,(368) & \textbf{35}\ \,(103) 
        & \textbf{505}\ \,(943) \\
        \midrule
        \multirow{3.4}{*}{}
        & \textit{w/(o) CtxBugs} 
        & \textbf{234}\ \,(278) & \textbf{350}\ \,(506) 
        & \textbf{238}\ \,(644) & \textbf{45}\ \,(128) 
        & \textbf{867}\ \,(1,556) \\
        \rowcolor{gray!25} \cellcolor{white} \multirow{-2}{*}{Qwen3}
        & \textit{w/ IsoBugs} 
        & \textbf{118}\ \,(150) & \textbf{230}\ \,(348) 
        & \textbf{101}\ \,(247) & \textbf{24}\ \,(72) 
        & \textbf{473}\ \,(817) \\
        \midrule
        \multirow{3.4}{*}{Kimi-K2}
        & \textit{w/(o) CtxBugs} 
        & \textbf{229}\ \,(280) & \textbf{354}\ \,(466) 
        & \textbf{280}\ \,(793) & \textbf{44}\ \,(122) 
        & \textbf{907}\ \,(1,661) \\
        \rowcolor{gray!25} \cellcolor{white} \multirow{-2}{*}{Kimi-K2}
        & \textit{w/ IsoBugs}
        & \textbf{39}\ \,(80) & \textbf{88}\ \,(165) 
        & \textbf{109}\ \,(324) & \textbf{25}\ \,(84) 
        & \textbf{261}\ \,(653) \\ 
        \bottomrule
    \end{tabular}
    \label{tab:dataset}
\end{table}

\subsection{Studied LLMs}
\label{sec:studied-llms}
To evaluate the performance of state-of-the-art LLMs in resolving \textit{CtxBugs}, we include four leading LLMs in our study. Since code adaptation requires both advanced natural language understanding and coding capabilities, our selection encompasses three general-purpose LLMs, including GPT-4o (2024-11-20)~\cite{OpenAI2024}, DeepSeek-V3 (2025-03-24)~\cite{DeepSeek2024} and Kimi-K2 (2025-07-11)~\cite{Kimi2025}, based on their exceptional results across various benchmarks, as well as one specialized code model, Qwen3-Coder-Plus (2025-07-22)~\cite{Qwen3Coder}, recognized for its extraordinary performance in coding tasks. An additional selection criterion is the accessibility of token generation probabilities, which allows us to analyze LLMs' confidence during generation. 

\subsection{Prompt Design}
In line with the effective prompt design established by Zhang et al.~\cite{Zhang2025IorI}, our adaptation prompt comprises three parts: (1) the user requirement, (2) the reused code for adaptation, and (3) the target class-level context. To simulate a realistic adaptation scenario, the targeted method itself and its callers are removed from this context. This structure provides LLMs with the essential contextual information required for successful code adaptation. Prompt templates and concrete examples are available in our replication package.

\subsection{Evaluation Metrics}
Simiar to our \textit{CtxBug} identification, we adopt both execution-based  and syntactic-based metrics to evaluate the effectiveness of LLMs in resolving \textit{CtxBugs}.

\textbf{Pass@k} metric estimates the likelihood that an LLM produces a correct solution within $k$ attempts based on unit test execution. In our study, we set $k=1$ to assess the correctness of the LLM's output. This selection is motivated by both the high computational cost of generating multiple solutions and its alignment with developer behavior, as studies show practitioners typically consider only the first suggestion from AI coding assistants~\cite{Barke2023,Spiess2025}.

\textbf{Resolution Rate (RR)} measures whether each \textit{CtxBug} has been correctly resolved by performing an exact match against the reference solution. For each AST node in the solution code at the \textit{CtxBug} location, we find its corresponding node in the LLM-generated code using GumTree. A \textit{CtxBug} is considered resolved only if a matching node is found and its textual content is identical to the solution. This metric is applicable for evaluating LLMs' adaptation of code without \textit{CtxBugs} or with \textit{Isobugs}, as their locations for the required adaptations are consistent. 

\subsection{Implementation Details}
LLMs' different randomness settings can lead to different outcomes~\cite{Ouyang2023}. In our evaluation, we are interested in the most preferred solution generated by the LLMs without any scaling. Therefore, we set the temperature to 0 for all LLMs to eliminate the randomness. This configuration is also an established best practice when using the Pass@1 to measure only one generated solution~\cite{Chen2021,Liu2021,Spiess2025}. All experiments are conducted with four GeForce RTX 4090-24G GPUs on Ubuntu 22.04.

\section{Results}

\subsection{RQ1: Effectiveness of LLMs}

\begin{table}[htbp]
    \scriptsize
    \centering
    \caption{LLMs' performance in \textit{CtxBug} resolution. \textbf{Model-G} and \textbf{Model-T} represent the generator model and the tested model respectively.}
    \begin{tabular}{cl|C{22pt}C{22pt}C{22pt}C{22pt}C{22pt}C{22pt}C{22pt}C{22pt}|C{22pt}C{22pt}}
        \toprule
        \multirow{2.4}{*}{\textbf{Model-G}} 
        & \multirow{2.4}{*}{\textbf{Model-T}}
        & \multicolumn{2}{c}{\textbf{Interface}} 
        & \multicolumn{2}{c}{\textbf{Functionality}}
        & \multicolumn{2}{c}{\textbf{Identifier}}
        & \multicolumn{2}{c|}{\textbf{Dependency}}
        & \multicolumn{2}{c}{\textbf{All Tasks}} \\
        \cmidrule(lr){3-12}
        & & \textit{Pass@1} & \textit{RR} & \textit{Pass@1} & \textit{RR} & \textit{Pass@1} & \textit{RR} & \textit{Pass@1} & \textit{RR} & \textit{Pass@1} & \textit{RR} \\
        \midrule
        & GPT-4o
        & 67.18 & 67.81 & 33.77 & 21.93
        & 67.34 & \textbf{60.96} & 43.75 & 44.04 
        & 53.20 & \textbf{50.14} \\ \rowcolor{gray!25} \cellcolor{white} \multirow{-0.5}{*}{GPT-4o} 
        & DS-V3 
        & 66.02 & 67.81 & 34.04 & 23.94 
        & 66.33 & 58.68 & 47.92 & 41.28 
        & 52.90 & 49.46 \\ 
        & Qwen3 
        & \textbf{68.73} & \textbf{69.69} & \textbf{35.36} & 24.75 
        & 68.35 & 57.84 & 45.83 & 41.28 
        & \textbf{54.63} & 49.63 \\ 
        \rowcolor{gray!25} \cellcolor{white} & Kimi-K2 
        & 66.80 & 67.50 & \textbf{35.36} & \textbf{25.35} 
        & \textbf{69.02} & 58.44 & \textbf{52.08} & \textbf{44.95} 
        & \textbf{54.63} & 49.91 \\ 
        \midrule
        & GPT-4o 
        & 68.18 & 65.73 & \textbf{38.25} & \textbf{28.51} 
        & 69.00 & \textbf{62.93} & 44.68 & 56.00 & 55.40 & 52.87 \\ 
        \rowcolor{gray!25} \cellcolor{white} \multirow{-0.5}{*}{DS-V3} 
        & DS-V3  
        & 66.94 & 66.43 & 35.52 & 27.70 
        & 65.31 & 61.40 & 51.06 & 56.80 & 53.24 & 52.10 \\
        & Qwen3 
        & 68.60 & 65.03 & 36.34 & 28.31 
        & 70.48 & 61.78 & \textbf{61.70} & \textbf{61.60} 
        & 56.05 & 52.58 \\ 
        \rowcolor{gray!25} \cellcolor{white} & Kimi-K2 
        & \textbf{70.25} & \textbf{67.83} & 37.70 & 28.31 
        & \textbf{70.85} & 62.42 & 57.45 & 56.80 
        & \textbf{56.91} & \textbf{52.99} \\
        \midrule
        & GPT-4o 
        & 70.09 & 65.47 & 37.43 & 29.05 
        & \textbf{68.49} & \textbf{62.89} & 51.11 & 57.81
        & 55.48 & \textbf{51.93} \\ 
        \rowcolor{gray!25} \cellcolor{white} \multirow{-0.5}{*}{Qwen3} 
        & DS-V3 
        & 70.51 & 66.19 & 35.43 & 29.84 
        & 66.39 & 59.78 & 53.33 & 54.69 
        & 54.33 & 50.77 \\
        & Qwen3  
        & 65.81 & 62.59 & \textbf{39.43} & \textbf{31.62} 
        & 66.39 & 58.07 & 53.33 & \textbf{60.94} & 54.67 & 50.51 \\ 
        \rowcolor{gray!25} \cellcolor{white} & Kimi-K2 
        & \textbf{70.94} & \textbf{66.91} & 38.86 & 29.84 
        & 68.07 & 61.18 & \textbf{57.78} & 57.81
        & \textbf{56.52} & 51.74 \\
        \midrule
        & GPT-4o 
        & 68.12 & 67.50 & 33.62 & 22.32 
        & \textbf{72.86} & 68.98 & 55.27 & 53.28
        & 55.35 & 54.49 \\
        \rowcolor{gray!25} \cellcolor{white} \multirow{-0.5}{*}{Kimi-K2} 
        & DS-V3 
        & \textbf{71.18} & 68.21 & 33.62 & 22.75 
        & 72.50 & 68.60 & 52.27 & 54.10
        & 56.01 & 54.61 \\
        & Qwen3  
        & 68.12 & 66.43 & \textbf{35.59} & \textbf{24.03} 
        & \textbf{72.86} & \textbf{69.23} & \textbf{65.91} & \textbf{60.66} 
        & \textbf{56.78} & \textbf{55.45} \\ 
        \rowcolor{gray!25} \cellcolor{white} & Kimi-K2 
        & 69.87 & \textbf{69.64} & 33.90 & 22.75
        & \textbf{72.86} & 68.98 & 50.00 & 58.20 & 55.79 & 55.33 \\
        \midrule
        & GPT-4o 
        & 68.36 & 66.67 & 35.75 & 25.51 & 69.43 & \textbf{63.95} & 48.54 & 53.10 & 54.82 & 52.33 \\
        \rowcolor{gray!25} \cellcolor{white} \multirow{-0.5}{*}{\textbf{Average}} 
        & DS-V3 
        & 68.57 & 67.18 & 34.65 & 26.12 & 67.68 & 62.18 & 51.09 & 52.07 & 54.09 & 51.72 \\
        & Qwen3  
        & 67.84 & 66.07 & \textbf{36.65} & \textbf{27.24} & 69.61 & 61.85 & \textbf{56.52} & \textbf{56.61} & 55.53 & 52.03 \\ 
        \rowcolor{gray!25} \cellcolor{white} &  Kimi-K2
        & \textbf{69.40} & \textbf{67.95} & 36.44 & 26.63 & \textbf{70.26} & 62.77 & 54.35 & 54.75 & \textbf{55.93} & \textbf{52.47} \\
        \bottomrule
    \end{tabular}
    \label{tab:res-rq1}
\end{table}

\begin{wrapfigure}[10]{R}{0.4\textwidth}
    \centering
    \includegraphics[width=0.38\columnwidth]{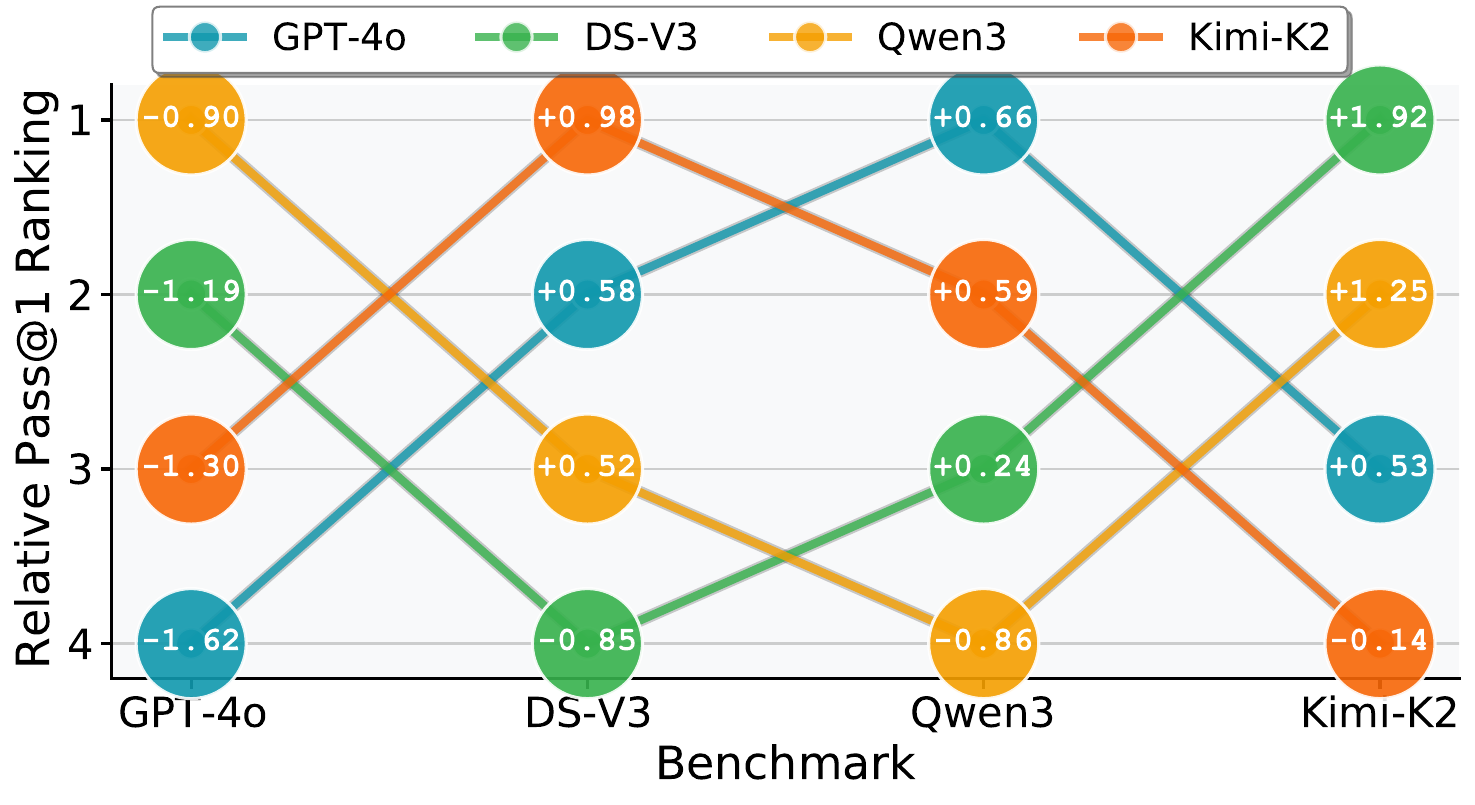}
    \caption{Relative \textit{Pass@1} ranking of LLMs.}
    \label{fig:res-rank}
\end{wrapfigure}

Table~\ref{tab:res-rq1} presents the performance of each LLM in resolving \textit{CtxBugs} on our benchmark constructed by \textit{CtxBugGen}. To ensure a fair evaluation, each LLM is tested on \textit{CtxBugs} generated by all four models.
We also report their overall average performance weighted by benchmark size. Kimi-K2 achieves the best performance among all LLMs, with a Pass@1 of 55.93\% and an RR of 52.47\% respectively. 
Across four adaptation tasks, LLMs perform better at resolving \textit{Interface Specification} and \textit{Identifier Reference} \textit{CtxBugs}. This is likely because these tasks are more syntax-aware, allowing models to rely on textual patterns.
In contrast, \textit{Functionality Customization} proves most challenging, with all LLMs achieving under 40\% Pass@1 and under 30\% RR. The reason is that this task requires deep understanding of code semantics and accurate identification of contextual constraints in both code and natural language.
Regarding \textit{CtxBugs} in the \textit{Dependency Constraint} task, Qwen3-Coder-Plus exhibits a particular strength, achieving over 60\% on both Pass@1 and RR for benchmarks generated by DeepSeek-V3 and Kimi-K2.
Besides, the gap between Pass@1 and RR is more pronounced for \textit{Functionality Customization} and \textit{Identifier Reference}. Since their cases may contain multiple interdependent \textit{CtxBugs} (introduced by Rule 3, 4, 8 \& 9), failing to identify the complete set prevents LLMs from resolving any of them, leading to a severe drop in RR.

Besides, we evaluate differences of LLMs' performance across \textit{CtxBugs} generated by different LLMs. Considering their inherent differences in capabilities, we compute each LLM's relative Pass@1 and its corresponding ranking on each benchmark, as shown in Fig.~\ref{fig:res-rank}. A key finding is that all four LLMs rank last on the data generated by themselves, which is statistically significant ($p=0.0039$ via Fisher's Sign Test~\cite{Hollander2013}). This suggests that LLMs perceive their own generated tokens as more plausible, making them more susceptible to such \textit{CtxBugs}.
We also observe a symmetrical performance pattern across model pairs, e.g., GPT-4o performs best on data generated by Qwen3, and vice versa. This implies that their biases on semantic correctness are also symmetrical.

\begin{center}
    \begin{tcolorbox}[colback=lgray,colframe=black,width=\linewidth,arc=1mm,boxrule=1pt,top=2pt,bottom=2pt,left=3pt,right=3pt]
        \textbf{Finding 1:} LLMs can perform correct adaptations in up to 55.93\% cases and resolve at most 52.47\% \textit{CtxBugs}. \textit{CtxBugs} in \textit{Functionality Customization} are the most difficult to resolve, with a RR of 27.24\%. Besides, LLMs are less effective in resolving \textit{CtxBugs} generated by themselves.
    \end{tcolorbox}
\end{center}

\subsection{RQ2: Impact of \textit{CtxBugs}}

\begin{table}[htbp]
    \tiny
    \centering
    \caption{Comparison of LLMs' Performance in Adapting \textit{Code w/ CtxBugs} and \textit{Code w/o CtxBugs}.}
    \begin{tabular}{cc|C{22pt}C{22pt}C{22pt}C{22pt}C{22pt}C{22pt}C{22pt}C{22pt}|C{22pt}C{22pt}}
        \toprule
        \multirow{2.4}{*}{\textbf{Model}} 
        & \multirow{2.4}{*}{\textbf{Setting}}
        & \multicolumn{2}{c}{\textbf{Interface}} 
        & \multicolumn{2}{c}{\textbf{Functionality}}
        & \multicolumn{2}{c}{\textbf{Identifier}}
        & \multicolumn{2}{c|}{\textbf{Dependency}}
        & \multicolumn{2}{c}{\textbf{All Tasks}} \\
        \cmidrule(lr){3-12}
        & & \textit{Pass@1} & \textit{RR} & \textit{Pass@1} & \textit{RR} & \textit{Pass@1} & \textit{RR} & \textit{Pass@1} & \textit{RR} & \textit{Pass@1} & \textit{RR} \\
        \midrule
        & \textit{w/ CtxBugs} 
        & 67.18 & 67.81 & 33.77 & 21.93 
        & 67.34 & 60.96 & 43.75 & 44.04 & 53.20 & 50.14 \\
        \rowcolor{gray!25} \cellcolor{white} \multirow{-0.7}{*}{GPT-4o}
        & \textit{w/o CtxBugs}
        & 74.52 & 76.56 & 53.03 & 46.68 
        & 83.50 & 84.43 & 68.75 & 67.89 & 68.67 & 71.32 \\
        \cmidrule(l){2-12}
        & \textit{Relative \%} 
        & \textcolor{textred}{\textbf{9.85\,$\downarrow$}} & \textcolor{textred}{\textbf{11.43\,$\downarrow$}} & \textcolor{textRed}{\textbf{36.32\,$\Downarrow$}} & \textcolor{textRed}{\textbf{53.02\,$\Downarrow$}} & \textcolor{textred}{\textbf{19.35\,$\downarrow$}} & \textcolor{textRed}{\textbf{27.80\,$\Downarrow$}} & \textcolor{textRed}{\textbf{36.36\,$\Downarrow$}} & \textcolor{textRed}{\textbf{35.13\,$\Downarrow$}} & \textcolor{textred}{\textbf{22.53\,$\downarrow$}} & \textcolor{textRed}{\textbf{29.70\,$\Downarrow$}} \\ 
        \midrule
        \multirow{3.4}{*}{DS-V3}
        & \textit{w/ CtxBugs}
        & 66.94 & 66.43 & 35.52 & 27.70 
        & 65.31 & 61.40 & 51.06 & 56.80 & 53.24 & 52.10 \\ 
        \rowcolor{gray!25} \cellcolor{white} \multirow{-0.7}{*}{DS-V3}
        & \textit{w/o CtxBugs}
        & 77.27 & 73.43 & 53.83 & 50.71 
        & 88.93 & 84.20 & 70.21 & 72.00 & 71.06 & 71.72 \\
        \cmidrule(l){2-12}
        & \textit{Relative \%} 
        & \textcolor{textred}{\textbf{13.37\,$\downarrow$}} & \textcolor{textred}{\textbf{9.53\,$\downarrow$}} & \textcolor{textRed}{\textbf{34.01\,$\Downarrow$}} & \textcolor{textRed}{\textbf{45.38\,$\Downarrow$}} & \textcolor{textRed}{\textbf{26.56\,$\Downarrow$}} & \textcolor{textRed}{\textbf{27.08\,$\Downarrow$}} & \textcolor{textRed}{\textbf{27.28\,$\Downarrow$}} & \textcolor{textred}{\textbf{21.11\,$\downarrow$}} & \textcolor{textRed}{\textbf{25.08\,$\Downarrow$}} & \textcolor{textRed}{\textbf{27.36\,$\Downarrow$}} \\  
        \midrule
        \multirow{3.4}{*}{Qwen3}
        & \textit{w/ CtxBugs}
        & 65.81 & 62.59 & 39.43 & 31.62 
        & 66.39 & 58.07 & 53.33 & 60.94 & 54.67 & 50.51 \\
        \rowcolor{gray!25} \cellcolor{white} \multirow{-0.7}{*}{Qwen3}
        & \textit{w/o CtxBugs} 
        & 77.78 & 72.66 & 58.29 & 54.94 
        & 86.55 & 82.76 & 68.89 & 78.91 & 71.86 & 71.59 \\ 
        \cmidrule(l){2-12}
        & \textit{Relative \%}
        & \textcolor{textred}{\textbf{15.39\,$\downarrow$}} & \textcolor{textred}{\textbf{13.86\,$\downarrow$}} & \textcolor{textRed}{\textbf{32.36\,$\Downarrow$}} & \textcolor{textRed}{\textbf{42.45\,$\Downarrow$}} & \textcolor{textred}{\textbf{23.29\,$\downarrow$}} & \textcolor{textRed}{\textbf{29.83\,$\Downarrow$}} & \textcolor{textred}{\textbf{22.59\,$\downarrow$}} & \textcolor{textred}{\textbf{22.77\,$\downarrow$}} & \textcolor{textred}{\textbf{23.92\,$\downarrow$}} & \textcolor{textRed}{\textbf{29.45\,$\Downarrow$}}\\ 
        \midrule
        \multirow{3.4}{*}{Kimi-K2}
        & \textit{w/ CtxBugs}
        & 69.87 & 69.64 & 33.90 & 22.75
        & 72.86 & 68.98 & 50.00 & 58.20 & 55.79 & 55.33 \\ 
        \rowcolor{gray!25} \cellcolor{white} \multirow{-0.7}{*}{Kimi-K2}
        & \textit{w/o CtxBugs}
        & 75.55 & 73.57 & 53.95 & 47.00  
        & 88.57 & 86.25 & 68.18 & 77.05 & 70.78 & 72.43 \\
        \cmidrule(l){2-12}
        & \textit{Relative \%}
        & \textcolor{textred}{\textbf{7.52\,$\downarrow$}} & \textcolor{textred}{\textbf{5.34\,$\downarrow$}} & \textcolor{textRed}{\textbf{37.16\,$\Downarrow$}} & \textcolor{textRed}{\textbf{51.60\,$\Downarrow$}} & \textcolor{textred}{\textbf{17.74\,$\downarrow$}} & \textcolor{textred}{\textbf{20.02\,$\downarrow$}} & \textcolor{textRed}{\textbf{26.66\,$\Downarrow$}} & \textcolor{textred}{\textbf{24.46\,$\downarrow$}} & \textcolor{textred}{\textbf{21.18\,$\downarrow$}} & \textcolor{textred}{\textbf{23.61\,$\downarrow$}} \\ 
        \midrule
        \textbf{Average} & \textit{Relative \%} 
        & \textcolor{textred}{\textbf{11.52\,$\downarrow$}} & \textcolor{textred}{\textbf{10.08\,$\downarrow$}} & \textcolor{textRed}{\textbf{34.99\,$\Downarrow$}} & \textcolor{textRed}{\textbf{48.04\,$\Downarrow$}} & \textcolor{textred}{\textbf{21.60\,$\downarrow$}} & \textcolor{textRed}{\textbf{26.03\,$\Downarrow$}} & \textcolor{textRed}{\textbf{28.35\,$\Downarrow$}} & \textcolor{textRed}{\textbf{25.55\,$\Downarrow$}} & \textcolor{textred}{\textbf{23.16\,$\downarrow$}} & \textcolor{textRed}{\textbf{27.53\,$\Downarrow$}} \\   
        \bottomrule
    \end{tabular}
    \label{tab:rq2-w/o-ctxbug}
\end{table}

Table~\ref{tab:rq2-w/o-ctxbug} compares LLMs' performance on adaptation with or without \textit{CtxBugs}. The presence of \textit{CtxBugs} causes an average decrease ratio of 23.16\% in Pass@1 and 27.53\% in RR, suggesting that LLMs are easily misled by \textit{CtxBugs} and perform worse than completing code from scratch. Across task types, \textit{Functionality Customization} shows the most severe degradation (34.99\% in Pass@1, 48.04\% in RR), while \textit{Interface Specification} is less affected. This is largely because that functionality-related \textit{CtxBugs} require deeper semantic reasoning, whereas interface violations often provide clearer syntactic clues for resolution. 

\begin{table}[htbp]
    \tiny
    \centering
    \caption{Comparison of LLMs' Performance in Adapting \textit{Code w/ CtxBugs} and \textit{Code w/ IsoBugs}.} 
    \begin{tabular}{cc|C{22pt}C{22pt}C{22pt}C{22pt}C{22pt}C{22pt}C{22pt}C{22pt}|C{22pt}C{22pt}}
        \toprule
        \multirow{2.4}{*}{\textbf{Model}} 
        & \multirow{2.4}{*}{\textbf{Setting}}
        & \multicolumn{2}{c}{\textbf{Interface}} 
        & \multicolumn{2}{c}{\textbf{Functionality}}
        & \multicolumn{2}{c}{\textbf{Identifier}}
        & \multicolumn{2}{c|}{\textbf{Dependency}}
        & \multicolumn{2}{c}{\textbf{All Tasks}} \\
        \cmidrule(lr){3-12}
        & & \textit{Pass@1} & \textit{RR} & \textit{Pass@1} & \textit{RR} & \textit{Pass@1} & \textit{RR} & \textit{Pass@1} & \textit{RR} & \textit{Pass@1} & \textit{RR} \\
        \midrule
        & \textit{w/ CtxBugs}
        & 59.82 & 49.69 & 37.46 & 23.65 
        & 61.06 & 57.80 & 46.67 & 45.28 & 49.39 & 44.96 \\
        \rowcolor{gray!25} \cellcolor{white} \multirow{-0.7}{*}{GPT-4o}
        & \textit{w/ IsoBugs} 
        & 60.71 & 58.39 & 47.77 & 46.53 
        & 65.87 & 74.27 & 48.89 & 64.15 & 55.79 & 62.63 \\ 
        \cmidrule(l){2-12}
        & \textit{Relative \%} 
        & \textcolor{textred}{\textbf{1.47\,$\downarrow$}} & \textcolor{textred}{\textbf{14.90\,$\downarrow$}} & \textcolor{textred}{\textbf{21.58\,$\downarrow$}} & \textcolor{textRed}{\textbf{49.17\,$\Downarrow$}} & \textcolor{textred}{\textbf{7.30\,$\downarrow$}} & \textcolor{textred}{\textbf{22.18\,$\downarrow$}} & \textcolor{textred}{\textbf{4.54\,$\downarrow$}} & \textcolor{textRed}{\textbf{29.42\,$\Downarrow$}} & \textcolor{textred}{\textbf{11.47\,$\downarrow$}} & \textcolor{textRed}{\textbf{28.21\,$\Downarrow$}} \\ 
        \midrule
        \multirow{3.4}{*}{DS-V3}
        & \textit{w/ CtxBugs}
        & 59.00 & 57.89 & 33.61 & 25.96 
        & 59.85 & 55.98 & 51.43 & 61.17 & 46.73 & 46.02 \\ 
        \rowcolor{gray!25} \cellcolor{white} \multirow{-0.7}{*}{DS-V3}
        & \textit{w/ IsoBugs} 
        & 55.00 & 57.14 & 37.82 & 41.59 
        & 65.91 & 75.27 & 68.57 & 81.55 & 50.69 & 61.29 \\
        \cmidrule(l){2-12}
        & \textit{Relative \%} 
        & \textcolor{textgreen}{\textbf{7.27\,$\uparrow$}} & \textcolor{textgreen}{\textbf{1.31\,$\uparrow$}} & \textcolor{textred}{\textbf{11.13\,$\downarrow$}} & \textcolor{textRed}{\textbf{37.58\,$\Downarrow$}} & \textcolor{textred}{\textbf{9.19\,$\downarrow$}} & \textcolor{textRed}{\textbf{25.63\,$\Downarrow$}} & \textcolor{textred}{\textbf{25.00\,$\downarrow$}} & \textcolor{textred}{\textbf{24.99\,$\downarrow$}} & \textcolor{textred}{\textbf{7.81\,$\downarrow$}} & \textcolor{textred}{\textbf{24.91\,$\downarrow$}} \\  
        \midrule
        & \textit{w/ CtxBugs}
        & 55.93 & 51.33 & 38.70 & 29.60 
        & 61.39 & 57.09 & 41.67 & 48.61 & 47.99 & 43.57 \\
        \rowcolor{gray!25} \cellcolor{white} \multirow{-0.7}{*}{Qwen3}
        & \textit{w/ IsoBugs} 
        & 55.08 & 56.67 & 44.78 & 50.57 
        & 72.28 & 78.13 & 41.67 & 65.28 & 53.07 & 61.32 \\
        \cmidrule(l){2-12}
        & \textit{Relative \%}
        & \textcolor{textgreen}{\textbf{1.54\,$\uparrow$}} & \textcolor{textred}{\textbf{9.42\,$\downarrow$}} & \textcolor{textred}{\textbf{13.58\,$\downarrow$}} & \textcolor{textRed}{\textbf{41.47\,$\Downarrow$}} & \textcolor{textred}{\textbf{15.07\,$\downarrow$}} & \textcolor{textRed}{\textbf{26.93\,$\Downarrow$}} & \textbf{0.00\,=} & \textcolor{textRed}{\textbf{25.54\,$\Downarrow$}} & \textcolor{textred}{\textbf{9.57\,$\downarrow$}} & \textcolor{textRed}{\textbf{28.95\,$\Downarrow$}} \\  
        \midrule
        \multirow{3.4}{*}{Kimi-K2}
        & \textit{w/ CtxBugs}
        & 48.72 & 61.25 & 30.68 & 21.21 
        & 71.56 & 69.44 & 48.00 & 60.71 & 52.11 & 55.13 \\ 
        \rowcolor{gray!25} \cellcolor{white} \multirow{-0.7}{*}{Kimi-K2}
        & \textit{w/ IsoBugs} 
        & 51.28 & 65.00 & 40.91 & 55.15 
        & 71.56 & 82.72 & 56.00 & 71.43 & 56.70 & 72.13 \\
        \cmidrule(l){2-12}
        & \textit{Relative \%}
        & \textcolor{textred}{\textbf{4.99\,$\downarrow$}} & \textcolor{textred}{\textbf{5.77\,$\downarrow$}} & \textcolor{textRed}{\textbf{25.01\,$\Downarrow$}} & \textcolor{textRed}{\textbf{61.54\,$\Downarrow$}} & \textbf{0.00\,=} & \textcolor{textred}{\textbf{16.05\,$\downarrow$}} & \textcolor{textred}{\textbf{14.29\,$\downarrow$}} & \textcolor{textred}{\textbf{15.01\,$\downarrow$}} & \textcolor{textred}{\textbf{8.10\,$\downarrow$}} & \textcolor{textred}{\textbf{23.57\,$\downarrow$}} \\  
        \midrule
        \textbf{Average} & \textit{Relative \%} 
        & \textcolor{textgreen}{\textbf{1.49\,$\uparrow$}} & \textcolor{textred}{\textbf{7.82\,$\downarrow$}} & \textcolor{textred}{\textbf{16.83\,$\downarrow$}} & \textcolor{textRed}{\textbf{45.49\,$\Downarrow$}} & \textcolor{textred}{\textbf{7.73\,$\downarrow$}} & \textcolor{textred}{\textbf{22.48\,$\downarrow$}} & \textcolor{textred}{\textbf{11.13\,$\downarrow$}} & \textcolor{textred}{\textbf{24.09\,$\downarrow$}} & \textcolor{textred}{\textbf{9.56\,$\downarrow$}} & \textcolor{textRed}{\textbf{26.69\,$\Downarrow$}} \\  
        \bottomrule
    \end{tabular}
    \label{tab:rq2-ctx-iso}
\end{table}

To ensure a fair comparison between LLMs' performance in resolving \textit{CtxBugs} and \textit{IsoBugs}, we only sample \textit{CtxBug} instances that also appear in the \textit{IsoBug} benchmark. Results are shown in Table~\ref{tab:rq2-ctx-iso}. The average decrease of Pass@1 is 9.56\%, while RR still drops by 26.69\%. This indicates that LLMs are better at fixing individual \textit{IsoBugs} than \textit{CtxBugs}, while their resolution for both remain incomplete.
The most significant gap occurs in \textit{Functionality Customization}, with an average decrease of 16.83\%. Interestingly, \textit{IsoBugs} in interface are found as challenging as \textit{CtxBugs}, with even lower Pass@1. The reason is that LLMs often introduce unused default parameters as \textit{IsoBugs}, which are easily overlooked since they lack functional effects. We also observe performance differences across LLMs. For instance, DeepSeek-V3 struggles with dependency-related \textit{CtxBugs}, while Kimi-K2 shows the smallest gap in identifiers.

\begin{figure*}
    \centering
    \includegraphics[width=\linewidth]{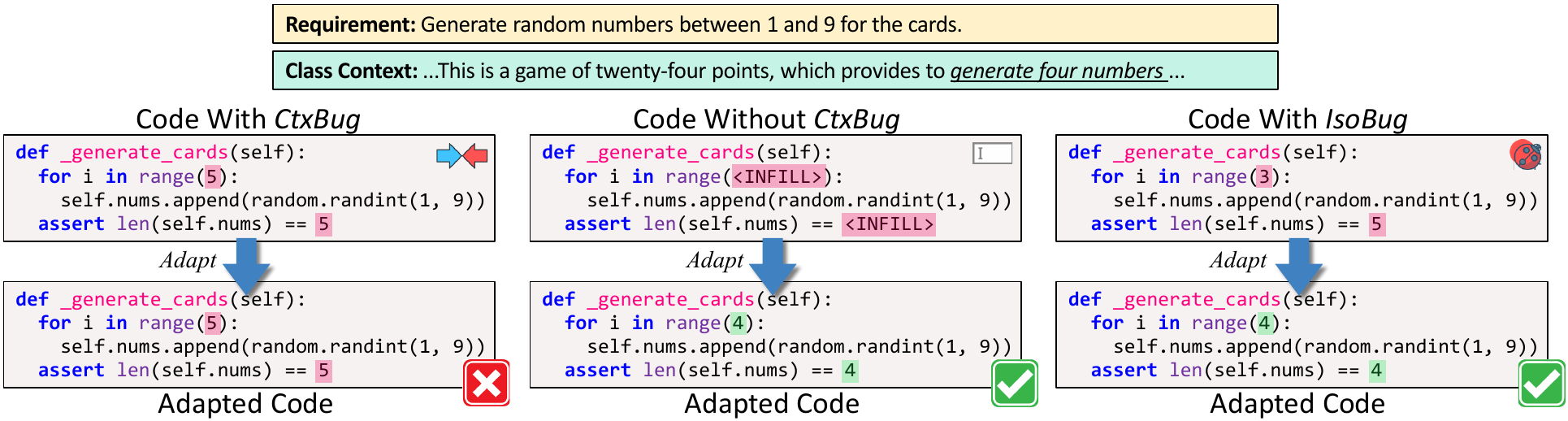}
    \caption{A failure example of adapting code with \textit{CtxBugs}.}
    \label{fig:rq2-example}
\end{figure*}

We further illustrate these scenarios with the example in Fig.~\ref{fig:rq2-example}. The task requires adapting a random number generation method for a Twenty-Four Point Game, which should generates exactly four numbers. Our framework introduces a \textit{CtxBug} that generates five numbers and includes an assertion for a length of five. In contrast, the \textit{CtxBug}-free code contains two placeholders to be filled, and the \textit{IsoBug} version incorrectly produces three numbers while asserting a length of five, creating an inherent inconsistency. The required adaptations are of similar complexity (i.e., two integers), ensuring a fair comparison. In this case, LLMs failed to resolve the \textit{CtxBug} but succeeded in the other two scenarios. This result indicates that the inconsistency of \textit{IsoBug} is locally detectable within the method, while the \textit{CtxBug} is locally correct but violates the game's requirement. LLMs still lack cross-context reasoning to identify such semantic mismatch during adaptation. 

\begin{center}
    \begin{tcolorbox}[colback=lgray,colframe=black,width=\linewidth,arc=1mm,boxrule=1pt,top=2pt,bottom=2pt,left=3pt,right=3pt]
        \textbf{Finding 2:} LLMs' adaptation performance is significantly impacted by \textit{CtxBugs}, with a degradation of 23.16\% and 27.53\% in Pass@1 and RR respectively. Besides, \textit{CtxBugs} are more difficult for LLMs to resolve than isolated errors.
    \end{tcolorbox}
\end{center}

\subsection{RQ3: Failure Analysis}
We employ stratified sampling to select erroneous adaptations that failed both test execution and syntactic differencing checks. The sample is drawn from the outputs of four evaluated LLMs across the whole benchmark (with 95\% CI, 5\% MOE). This process yields a set of 1,250 failed instances for manual analysis. Four annotators are invited to characterize each output from two dimensions: (1) the LLM's handling action for the \textit{CtxBug} and (2) the specific type of \textit{CtxBug} that caused failure in adaptation. The annotated data are available in the replication package.

For the first dimension, we enhance the taxonomy from Zhang et al.~\cite{Zhang2025IorI} to classify LLMs' handling actions into four categories, including:

\begin{itemize}[leftmargin=*]
    \item \textbf{Overlooked}: The LLM failed to identify the \textit{CtxBug} and replicated it in its output.
    \item \textbf{Invalid}: The LLM identified the \textit{CtxBug} but applied an incorrect adaptation.
    \item \textbf{Unexpected}: The LLM resolved the \textit{CtxBug} but also introduced new bugs elsewhere.
    \item \textbf{Others}: LLM's output is unrelated to the \textit{CtxBug} resolution, e.g., empty responses, reimplementations or code with major deletions.
\end{itemize}

\begin{figure}[htbp]
    \centering
    \subfigure[Distribution of LLMs' handling actions]{
        \includegraphics[width=0.51\linewidth]{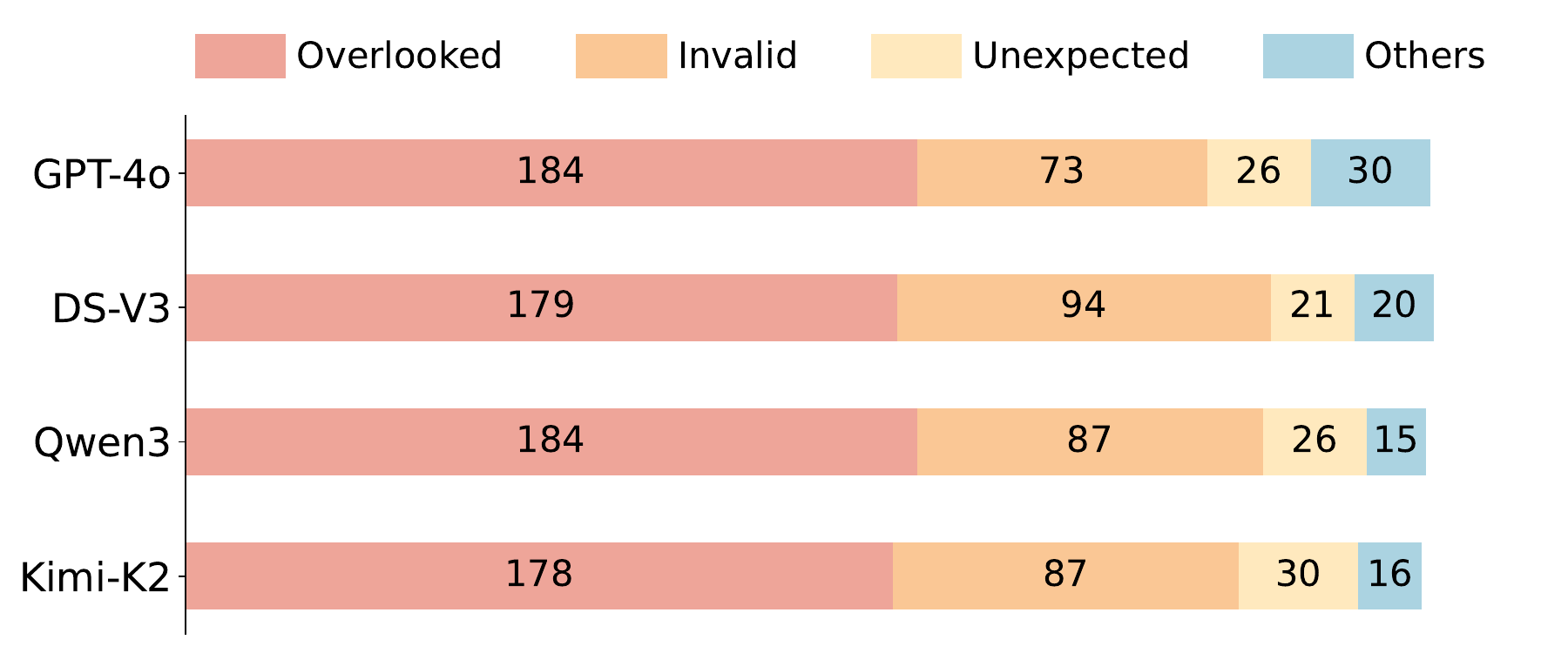}
        \label{fig:res-action}
    }
    \subfigure[\textit{Unexpected} handling action of \textit{CtxBug}]{
        \includegraphics[width=0.45\linewidth]{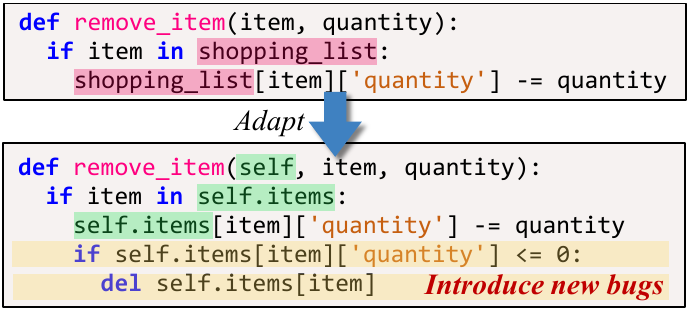}
        \label{fig:rq3-introduce}
    }
    \caption{LLMs' erroneous handling actions for \textit{CtxBugs}.}
    \label{fig:rq3-part1}
\end{figure}

\begin{table}[htbp]
    \scriptsize
    \centering
    \caption{The Average Token Probability (ATP) of LLMs' Overlooked \textit{CtxBugs}.}
    \begin{tabular}{cC{60pt}C{60pt}C{60pt}C{60pt}C{60pt}}
    \toprule
    \textbf{Model} & GPT-4o & DeepSeek-V3 & Qwen3-Coder-Plus & Kimi-K2 & \textbf{Average} \\
    \midrule
    \textbf{ATP} & 0.982 & 0.988 & 0.978 & 0.984 & \textbf{0.983} \\
    \bottomrule
    \end{tabular}
    \label{tab:res-confidence}
\end{table}

The distribution of LLMs' handling actions is summarized in Fig.~\ref{fig:res-action}. In nearly 60\% of cases, LLMs completely overlooked \textit{CtxBugs}, directly replicating the buggy code in their output. To further investigate this replication behavior, we calculate the Average Token Probability (ATP) for LLM-generated token sequences of the \textit{CtxBugs}. ATP measures the LLMs' intrinsic confidence in their generations~\cite{Spiess2025}. We observe that LLMs all exhibit high confidence with an average ATP of 0.983, indicating they consider the bug-inducing tokens to be correct. In up to 30\% of cases, LLMs performed adaptations that are insufficient to fully resolve the \textit{CtxBugs}. Besides, in up to 9.65\% of cases, LLMs' adaptations to fix the original \textit{CtxBug} resulted in introducing new bugs. As shown in Fig.~\ref{fig:rq3-introduce}, an LLM correctly replaced an undefined variable with a class attribute but then erroneously removed a key when its value is non-positive, which is a unspecified behavior. This suggests that LLMs sometimes inappropriately rely on patterns from their pretraining knowledge (e.g., a rule for cleaning up zero-value items) when the target context provides no explicit guidance.

\begin{table}[htbp]
    \tiny
    \centering
    \caption{The distribution of \textit{CtxBug} types that LLMs failed to resolve in different adaptation tasks.}
    \begin{tabular}{clL{40pt}ccccc}
        \toprule
        \multirow{2.4}{*}{\textbf{Task}} & \multicolumn{1}{c}{\multirow{2.4}{*}{\textbf{Bug Type}}} & \multicolumn{1}{c}{\multirow{2.4}{*}{\textbf{Associated Rules}}} & \multicolumn{4}{c}{\textbf{Model}} & \multirow{2.4}{*}{\textbf{Average}} \\
        \cmidrule(lr){4-7}
        & & & GPT-4o & DS-V3 & Qwen3 & Kimi-K2 & \\
        \midrule
        \multirow{2}{*}{\textbf{Interface}}
        & Incorrect Parameter & Rule 1
        & \cellcolor{rL3} 15 (22.73\%)  & \cellcolor{rL2} 28 (32.56\%)  & \cellcolor{rL2} 30 (30.93\%)  & \cellcolor{rL2} 38 (40.86\%)  & \cellcolor{rL2} 111 (32.46\%) \\
        & Incorrect Return & Rule 2
        & \cellcolor{rL1} 51 (77.27\%)  & \cellcolor{rL1} 58 (67.44\%)  & \cellcolor{rL1} 67 (69.07\%)  & \cellcolor{rL1} 55 (59.14\%)  & \cellcolor{rL1} 231 (67.54\%) \\
        \midrule
        \multirow{7}{*}{\textbf{Functionality}}
        & Incorrect Condition & Rule 3, 4, 6
        & \cellcolor{rL3} 51 (29.31\%)  & \cellcolor{rL2} 58 (32.22\%)  & \cellcolor{rL2} 56 (30.43\%)  & \cellcolor{rL3} 52 (29.55\%)  & \cellcolor{rL2} 217 (30.39\%) \\
        & Incorrect Argument & Rule 3, 7
        & \cellcolor{rL3} 25 (14.37\%)  & \cellcolor{rL3} 32 (17.78\%)  & \cellcolor{rL3} 23 (12.50\%)  & \cellcolor{rL3} 25 (14.20\%)  & \cellcolor{rL3} 105 (14.71\%) \\
        & Incorrect Initialization & Rule 3, 5
        & \cellcolor{rL3} 23 (13.22\%)  & \cellcolor{rL3} 38 (21.11\%)  & \cellcolor{rL3} 31 (16.85\%)  & \cellcolor{rL3} 26 (14.77\%)  & \cellcolor{rL3} 118 (16.53\%) \\
        & Incorrect Calculation & Rule 3, 4, 5
        & \cellcolor{rL3} 21 (12.07\%)  & \cellcolor{rL3} 21 (11.67\%)  & \cellcolor{rL3} 34 (18.48\%)  & \cellcolor{rL3} 27 (15.34\%)  & \cellcolor{rL3} 103 (14.43\%) \\
        & Incorrect Data Operation & Rule 5, 7
        & \cellcolor{rL3} 29 (16.67\%)  & \cellcolor{rL3} 19 (10.56\%)  & \cellcolor{rL3} 31 (16.85\%)  & \cellcolor{rL3} 35 (19.89\%)  & \cellcolor{rL3} 114 (15.97\%) \\
        & Incorrect Type Operation & Rule 5, 7
        &  11 (6.32\%)  &  4 (2.22\%)  &  1 (0.54\%)  &  6 (3.41\%)  &  22 (3.08\%) \\
        & Incorrect Indexing & Rule 3
        &  14 (8.05\%)  &  8 (4.44\%)  &  8 (4.35\%)  &  5 (2.84\%)  &  35 (4.90\%) \\
        \midrule
        \multirow{3}{*}{\textbf{Identifier}}
        & Misassigned Identifier & Rule 9
        & \cellcolor{rL3} 10 (17.54\%)  & \cellcolor{rL3} 9 (15.00\%)  & \cellcolor{rL3} 7 (14.00\%)  & \cellcolor{rL3} 7 (12.07\%)  & \cellcolor{rL3} 33 (14.67\%) \\
        & Missing Context & Rule 8
        & \cellcolor{rL1} 36 (63.16\%)  & \cellcolor{rL2} 29 (48.33\%)  & \cellcolor{rL1} 28 (56.00\%)  & \cellcolor{rL2} 28 (48.28\%)  & \cellcolor{rL1} 121 (53.78\%) \\
        & Context Misuse & Rule 8
        & \cellcolor{rL3} 11 (19.30\%)  & \cellcolor{rL2} 22 (36.67\%)  & \cellcolor{rL3} 15 (30.00\%)  & \cellcolor{rL2} 23 (39.66\%)  & \cellcolor{rL2} 71 (31.56\%) \\
        \midrule
        \multirow{2}{*}{\textbf{Dependency}}
        & Missing Library & Rule 10
        & \cellcolor{rL1} 18 (60.00\%)  & \cellcolor{rL1} 18 (58.06\%)  & \cellcolor{rL1} 14 (56.00\%)  & \cellcolor{rL1} 9 (64.29\%)  & \cellcolor{rL1} 59 (59.00\%) \\
        & Library Misuse & Rule 10
        & \cellcolor{rL2} 12 (40.00\%)  & \cellcolor{rL2} 13 (41.94\%)  & \cellcolor{rL2} 11 (44.00\%)  & \cellcolor{rL2} 5 (35.71\%)  & \cellcolor{rL2} 41 (41.00\%) \\
        \bottomrule
    \end{tabular}
    \label{tab:res-rq3}
\end{table}

\begin{figure}[htbp]
    \centering
    \subfigure[\textit{CtxBug} in the return value]{
        \includegraphics[width=0.46\linewidth]{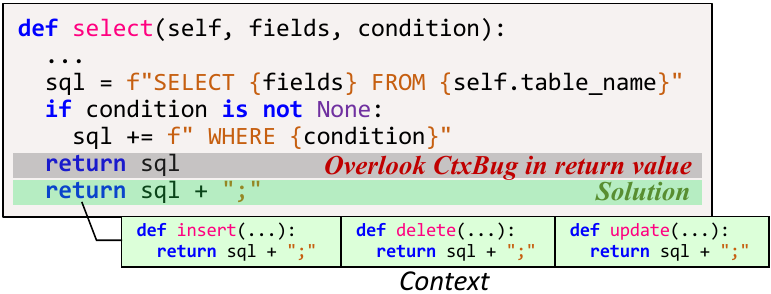}
        \label{fig:rq3-return}
    }
    \subfigure[\textit{CtxBug} in the conditional]{
        \includegraphics[width=0.50\linewidth]{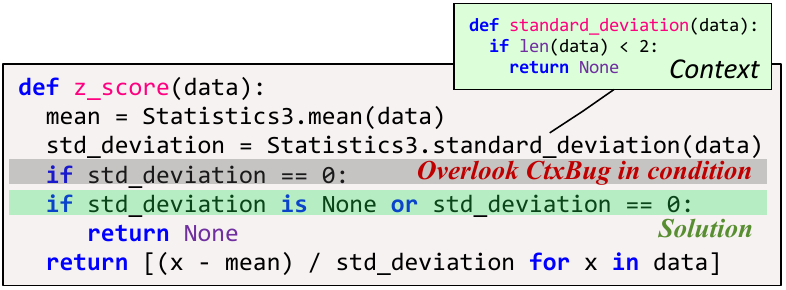}
        \label{fig:rq3-condition}
    }
    
    \subfigure[\textit{CtxBug} in the bounded identifiers]{
        \includegraphics[width=0.59\linewidth]{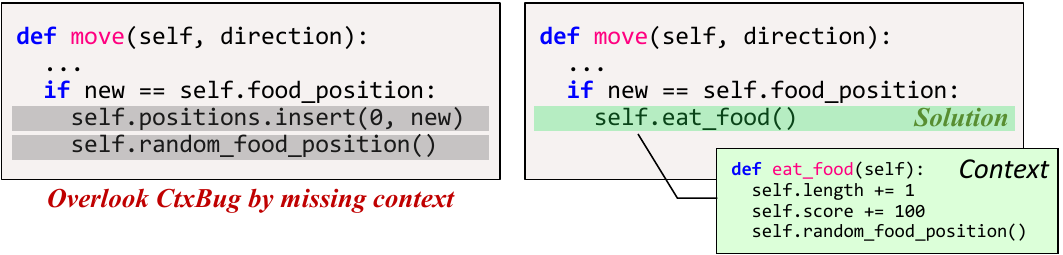}
        \label{fig:rq3-context}
    }
    \subfigure[\textit{CtxBug} in the library usage]{
        \includegraphics[width=0.37\linewidth]{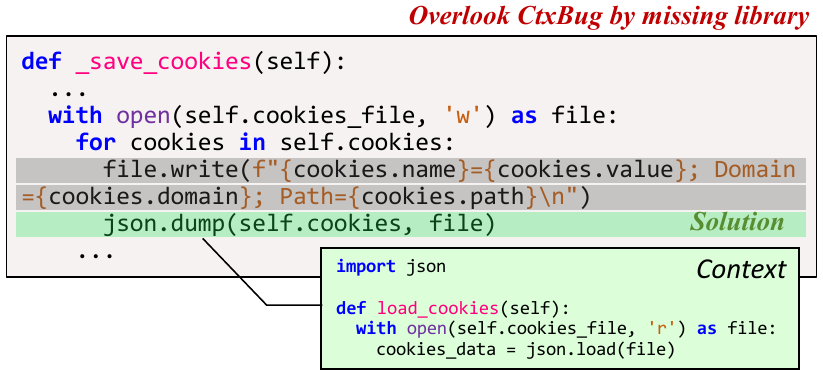}
        \label{fig:rq3-library-new}
    }
    \caption{Examples of LLMs' failures in adapting code with \textit{CtxBugs}.}
    \label{fig:rq3-failure-examples}
\end{figure}

Table~\ref{tab:res-rq3} summarizes the type distribution of \textit{CtxBugs} that LLMs failed to resolve. Since \textit{CtxBugs} are generated for specific adaptation tasks, our taxonomy and analysis are correspondingly scoped to these tasks individually. The detailed definition of each type is available in our replication package. 
\textit{Incorrect Return} is the most prevalent \textit{CtxBug} type in the \textit{Interface Specification} task. LLMs often failed to identify the inconsistencies between the outputs of the reused code and its context. A representative example is shown in Fig.\ref{fig:rq3-return}, where Kimi-K2 overlooked the missing semicolon in a return value, even though the context provided clear examples of the correct formatting in sibling methods.
\textit{CtxBugs} in conditions are the most challenging to resolve in the \textit{Functionality Customization}, with a proportion of about 30\% across all LLMs. As shown in Fig.\ref{fig:rq3-condition}, Qwen3 accepted the plausible but contextually incomplete condition (e.g., checking if a standard deviation is zero) instead of performing the necessary cross-context reasoning on the callee's output, whose edge case returns \textit{None}.
For identifiers, LLMs often ignore the pre-existing implementations to resolve \textit{CtxBugs}. For example, in a Snake game (Fig.\ref{fig:rq3-context}), Kimi-K2 failed to use the ``\textit{eat\_food()}'' method in the context (which handles the snake length and score increment) and instead copied the \textit{CtxBug} that merely added coordinates to the snake's position.
Finally, for dependency-related bugs, LLMs frequently missed using a required library or misused its API. 
Figure\ref{fig:rq3-library-new} illustrates a \textit{CtxBug} where GPT-4o failed to use the provided library \textit{json} in its implementation. The provided context ``\textit{load\_cookies}'' reads persisted cookie data in the JSON format with the required library. However, GPT-4o overlooked this dependency and maintained the incompatible data serialization in a line-by-line string format instead.

To investigate why LLMs overlooked \textit{CtxBugs}, we further prompt them to explain their reasoning alongside their code adaptations\footnote{The explanation prompt is provided in our replication package.}. We manually analyze the outputs of 50 samples to determine whether the failures originate from LLMs' inability to identify \textit{CtxBugs} or their inability to perform the repair.
Our results show that in 98\% of cases, LLMs failed to identify \textit{CtxBugs} entirely. 
For instance, in the case shown in Fig.\ref{fig:rq3-condition}, the LLM claimed ``\textit{No adaptations have been made because the provided code is correct...}'' In the remaining one case, LLMs successfully identified the \textit{CtxBug} and performed correct adaptations, likely due to the Chain-of-Thought (CoT) effects~\cite{Wei2022} induced by the explanation.

\begin{center}
    \begin{tcolorbox}[colback=lgray,colframe=black,width=\linewidth,arc=1mm,boxrule=1pt,top=2pt,bottom=2pt,left=3pt,right=3pt]
        \textbf{Finding 3:} In nearly 60\% of cases, LLMs overlook \textit{CtxBugs} and replicate them as correct code. Their weaknesses lie in resolving \textit{CtxBugs} related to return values, conditionals, and the incorporation of pre-existing implementations.
    \end{tcolorbox}
\end{center}

\section{Implication}
\textbf{Mind \textit{CtxBugs} and Consider Cross-Context Semantics in Code Reuse.} 
While fixing standalone bugs has long been a focus in software engineering~\cite{Xia2023}, our study emphasizes that resolving \textit{CtxBugs} is equally critical and often more challenging due to their semantic subtleties. Developers should carefully examine whether reused code aligns with contextual constraints. To support this need, we propose \textit{CtxBugGen} and a dedicated benchmark for systematically evaluating \textit{CtxBug} resolution, providing a foundation for future research and tool development in this domain.

\textbf{Double Check When Using LLMs for Code Adaptation.}
Our findings reveal that LLMs often overlook and replicate \textit{CtxBugs}, generating code that is locally plausible but contextually incompatible. Practitioners using LLMs for adaptation should therefore employ rigorous quality assurance, to validate their outputs. It is particularly necessary for functionality interacting with contextual elements, e.g., pre-existing methods, to avoid introducing subtle semantic errors.

\textbf{Develop Methods to Enhance Cross-Context Reasoning.}
Our study suggests LLMs currently lack of robust cross-context reasoning, i.e., they overly rely on local code correctness and fail to consider broader contextual constraints. Future work should explore: (1) \textit{Fine-tuning} on cross-context tasks like adaptation; (2) \textit{Prompting strategies} that explicitly guide contextual reasoning through chain-of-thoughts or self-reflection; and (3) \textit{Hybrid techniques} combining LLMs with static analysis to enhance their contextual awareness. Advancing these areas is essential for LLMs application in code adaptation tasks.

\section{Threats to Validity}
\textbf{Threats to Internal Validity.} 
Threats may arise from our adaptation task selection and benchmark construction. Although we derive four key tasks from literature review, they may not cover all adaptation scenarios (e.g., optimizations for efficiency). Future work should evaluate LLMs on a broader range. For benchmark construction, we use LLMs to generate both \textit{CtxBugs} and \textit{IsoBugs}, which may affect data quality (e.g., introduces randomness or model-specific noises and may not represent their real-world complexity). To mitigate this, we conduct a rigorous identification combining test execution and syntactic differencing. Our manual validation further confirms the quality of LLM-implanted bugs, which achieves high validity and strong inter-annotator agreements. Additionally, Table~\ref{tab:res-validated} reports LLMs' performance on our manually validated \textit{CtxBugs} subset. These results align closely with the full benchmark, confirming the reliability of our findings. Regarding the potential data leakage risk, we address it by applying code obfuscation techniques during benchmark construction and evaluation.

\begin{table}[htbp]
    \scriptsize
    \centering
    \caption{LLMs' performance on the full LLM-generated benchmark and our validated \textit{CtxBugs}.}
    \begin{tabular}{cl|C{22pt}C{22pt}C{22pt}C{22pt}C{22pt}C{22pt}C{22pt}C{22pt}|C{22pt}C{22pt}}
        \toprule
        \multirow{2.4}{*}{\textbf{Data}} 
        & \multirow{2.4}{*}{\textbf{Model-T}}
        & \multicolumn{2}{c}{\textbf{Interface}} 
        & \multicolumn{2}{c}{\textbf{Functionality}}
        & \multicolumn{2}{c}{\textbf{Identifier}}
        & \multicolumn{2}{c|}{\textbf{Dependency}}
        & \multicolumn{2}{c}{\textbf{All Tasks}} \\
        \cmidrule(lr){3-12}
        & & \textit{Pass@1} & \textit{RR} & \textit{Pass@1} & \textit{RR} & \textit{Pass@1} & \textit{RR} & \textit{Pass@1} & \textit{RR} & \textit{Pass@1} & \textit{RR} \\
        \midrule
        & GPT-4o 
        & 68.36 & 66.67 & 35.75 & 25.51 & 69.43 & \textbf{63.95} & 48.54 & 53.10 & 54.82 & 52.33 \\
        \rowcolor{gray!25} \cellcolor{white} \multirow{-0.5}{*}{Full} 
        & DS-V3 
        & 68.57 & 67.18 & 34.65 & 26.12 & 67.68 & 62.18 & 51.09 & 52.07 & 54.09 & 51.72 \\
        & Qwen3  
        & 67.84 & 66.07 & \textbf{36.65} & \textbf{27.24} & 69.61 & 61.85 & \textbf{56.52} & \textbf{56.61} & 55.53 & 52.03 \\ 
        \rowcolor{gray!25} \cellcolor{white} &  Kimi-K2
        & \textbf{69.40} & \textbf{67.95} & 36.44 & 26.63 & \textbf{70.26} & 62.77 & 54.35 & 54.75 & \textbf{55.93} & \textbf{52.47} \\
        \midrule
        & GPT-4o 
        & 72.53 & 71.19 & 34.78 & 26.25 
        & \textbf{75.53} & \textbf{66.89} & 50.00 & 45.45
        & \textbf{58.54} & \textbf{55.90} \\
        \rowcolor{gray!25} \cellcolor{white} \multirow{-0.5}{*}{Validated} 
        & DS-V3 
        & 67.03 & 66.95 & \textbf{37.39} & \textbf{29.38} 
        & 74.47 & 66.22 & 50.00 & 48.48
        & 57.59 & 55.74 \\
        & Qwen3  
        & 68.13 & 67.80 & 36.52 & 28.12 
        & 73.40 & 60.87 & \textbf{56.25} & \textbf{51.52} 
        & 57.59 & 53.11 \\ 
        \rowcolor{gray!25} \cellcolor{white} & Kimi-K2 
        & \textbf{73.63} & \textbf{72.03} & 33.91 & 28.12
        & \textbf{75.53} & 66.56 & 37.50 & 36.36 & 57.91 & \textbf{55.90} \\
        \bottomrule
    \end{tabular}
    \label{tab:res-validated}
\end{table}

\textbf{Threats to External Validity.}
The generalizability of our findings may be limited by the selection of LLMs and base benchmarks. While we evaluate four state-of-the-art LLMs based on their performance, accessibility, and specialty to highlight common bottlenecks, our results may not apply to other untest or emerging models.
Another limitation arises from our reliance on ClassEval as the base benchmark. Since ClassEval is restricted to Python and focuses exclusively on class-level contexts, our findings may not generalize to other languages (e.g., Java, Rust) or broader contexts (e.g., system-level adaptations). However, our proposed \textit{CtxBugGen} framework is designed to be language-agnostic. It can be adapted to these scenarios by integrating language-specific syntactic parsing and perturbation rules. In our future work, we plan to extend \textit{CtxBugGen} to multi-task, multi-language and multi-context scenarios.

\section{Conclusion}
This paper proposes \textit{CtxBugGen}, the first framework for generating \textit{CtxBugs} to evaluate LLMs on code adaptation tasks. 
Our empirical study with four state-of-the-art LLMs reveals their unsatisfactory performance in resolving \textit{CtxBugs}.
A key finding is that LLMs frequently overlook these bugs and replicate them in their outputs, suggesting an over-reliance on local code correctness while ignoring contextual constraints. Future work should extend beyond training and evaluating LLMs on standalone coding tasks and prioritize enhancing their cross-context reasoning abilities, which is critical for their reliable application in real-world code reuse and adaptation.

\section{Data Availability}
The replication package including all the data (benchmarks, human annotation for literature review, benchmark validation and failure analysis), code (for benchmark construction and empirical evaluation) and related materials (prompts, intermediate results, etc.) is available here\footnote{\url{https://github.com/ztwater/CtxBugGen}.}. The detailed guides are included in the README.md file under each folder. 

\begin{acks}
We thank the anonymous reviewers for their constructive feedback to improve our paper. We also gratefully acknowledge the support from the National Key Research and Development Program of China (Grant No.2023YFB4503802) and the National Natural Science Foundation of China (Grant No.62302515, No.62172426, and No.62332005).
\end{acks}

\bibliographystyle{ACM-Reference-Format}
\bibliography{ref}


\end{document}